\documentclass[aip,amsmath,amssymb,reprint]{revtex4-1}

\usepackage{color}
\usepackage{dcolumn}
\usepackage{graphicx}
\usepackage{subfigure}
\usepackage{amsmath}
\usepackage{comment}
\usepackage{hyperref} 
\graphicspath{{images_pdf/} } 

\usepackage{xr-hyper}
\usepackage{todonotes}

\newcommand{\bxi}{\boldsymbol{\xi}}

\newcommand{\bx}{{{\bf x}}}
\newcommand{\bff}{{{\bf f}}}
\newcommand{\bF}{{{\bf F}}}
\newcommand{\bT}{{{\bf T}}}

\newcommand{\bJ}{{\bf J}}
\newcommand{\bI}{{\bf I}}
\newcommand{\bu}{{\bf u}}
\newcommand{\bv}{{\bf v}}

\newcommand{\bpsi}{{\boldsymbol{\psi}}}
\newcommand{\bmu}{\boldsymbol{\mu}}
\newcommand{\bphi}{\boldsymbol{\phi}}
\newcommand{\bSigma}{\boldsymbol{\Sigma}}

\newcommand{\be}{\begin{equation}}
\newcommand{\ee}{\end{equation}}
\newcommand{\bea}{\begin{eqnarray}}
\newcommand{\eea}{\end{eqnarray}}

\newcommand{\qt}{{q}}

\newcommand{\bra}{{\langle}}
\newcommand{\ket}{{\rangle}}

\newcommand{\D}{{\mathcal D}}

\newcommand{\bC}{{\bf C}}
\newcommand{\bM}{{\mathcal M}}

\usepackage{orcidlink}

\makeatletter
\def\@email#1#2{%
 \endgroup
 \patchcmd{\titleblock@produce}
  {\frontmatter@RRAPformat}
  {\frontmatter@RRAPformat{\produce@RRAP{*#1\href{mailto:#2}{#2}}}\frontmatter@RRAPformat}
  {}{}
}
\makeatother

\begin{document}
	
\title{Dynamically selected steady states and criticality in non-reciprocal networks}
\author{Carles Martorell}
\affiliation{Departamento de Electromagnetismo y F{\'\i}sica de la Materia and Instituto Carlos I
de F{\'\i}sica Te{\'o}rica y Computacional, Universidad de Granada, E-18071, Granada, Spain}
\author{Rub\'en Calvo} 
\affiliation{Departamento de Electromagnetismo y F{\'\i}sica de la Materia and Instituto Carlos I
de F{\'\i}sica Te{\'o}rica y Computacional, Universidad de Granada, E-18071, Granada, Spain}
\author{Alessia Annibale} 
\affiliation{Department of Mathematics, King’s College London, SE11 6NJ London, United Kingdom}
\author{Miguel A. Mu\~noz} 
\affiliation{Departamento de Electromagnetismo y F{\'\i}sica de la Materia and Instituto Carlos I
de F{\'\i}sica Te{\'o}rica y Computacional, Universidad de Granada, E-18071, Granada, Spain}

\begin{abstract}
 We consider a simple neural network model, evolving via non-linear coupled stochastic differential equations, where neural couplings are random Gaussian variables with non-zero mean and arbitrary degree of reciprocity. Using a path-integral approach, we analyze the dynamics, averaged over the network ensemble, in the thermodynamic limit.
 Our results show that for any degree of reciprocity in the couplings, two types of criticality emerge, corresponding to ferromagnetic and spin-glass order, respectively. The critical lines separating the disordered from the  ordered phases is consistent with spectral properties of the coupling matrix, as derived from random matrix theory. Both ordered phases shrink as the non-reciprocity (i.e. asymmetry) in the couplings increases, with the spin-glass phase disappearing when the couplings become anti-symmetric. Non-fixed point steady state solutions are analysed for uncorrelated interactions. For such solutions the correlation function evolves according to a gradient-descent dynamics on a potential, which depends on the stationary variance. Our analysis shows that in the spin-glass region, the variance dynamically selected by the system leads the correlation function to evolve on the separatrix curve, limiting different realizable steady states, whereas in the ferromagnetic region, a fixed point solution is selected as the only realizable steady state. Solutions in the spin-glass region are unstable against perturbations that break time-translation invariance, indicating that in large single network instances, the motion is chaotic. Numerical analysis of Lyapunov exponents confirms 
 that chaotic behaviour emerges in the whole spin-glass region, for any value of the coupling correlations. While negative correlations increase the strength of chaos, positive ones reduce it, with chaos disappearing for reciprocal (i.e. symmetric) couplings, where marginal stability is attained. On the other hand, in finite size non-reciprocal networks, fixed points and limit cycles can arise in the spin-glass region, especially close to the critical line. Finally, we show that when the  strength of external noise exceeds a certain threshold, chaos is suppressed. Intriguing analogies between chaotic phases in non-equilibrium systems and spin-glass phases in equilibrium are put forward.
 \end{abstract}

	\maketitle

\begin{quotation}
Diverse equilibrium systems with heterogeneous interactions lie at the edge of stability.  
Such marginally stable states are dynamically selected as the most abundant ones or as those with the largest basins of attraction. On the other hand, systems with non-reciprocal (or asymmetric) interactions are inherently out of equilibrium, and exhibit a rich variety of steady states, including fixed points, limit cycles and chaotic trajectories. How are steady states dynamically selected away from equilibrium?  We address this question in a simple neural network model, with a tunable level of non-reciprocity. Our study reveals different types of ordered phases 
and it shows how non-equilibrium steady states are selected in each phase. 
In the spin-glass region, the system exhibits marginally stable behaviour for reciprocal (or symmetric) interactions and it smoothly transitions to chaotic dynamics, as the non-reciprocity (or asymmetry) in the couplings increases. Such region, on the other hand, shrinks and eventually disappears when couplings become anti-symmetric. Our results are relevant to advance the knowledge of disordered systems beyond the paradigm of reciprocal couplings, and to develop an interface between statistical physics of equilibrium spin-glasses and dynamical systems theory. 
\end{quotation}

\section{Introduction}
Countless systems, both natural and human-made ---such as species in ecosystems, neurons in neural networks, agents in financial markets and atomic magnetic moments in disordered materials--- can be visualized as heterogenous networks of interacting elements. In the case of fully connected systems with symmetric or reciprocal interactions, the general phenomenology is that upon increasing the heterogeneity in pairwise interactions, there is a transition from a phase with a single equilibrium to a ``spin-glass" phase with multiple equilibria \cite{SG-beyond,pedestrians,Nishimori,SG-2023}. The latter is characterized by ergodicity breaking, slow dynamics, memory, and the breakdown of time-translation invariance (TTI) among other non-trivial features. The spin-glass (SG) phase, is characterized by a complex energy landscape with many minima, separated by large barriers, hence different copies or "\emph{replicas}" of the system typically relax in distinct regions of the phase space \cite{SG-beyond}. This leads to the phenomenon of replica-symmetry breaking (RSB), namely the breakdown of symmetry among different replicas, when averaging over the distribution of interactions \cite{SG-beyond}. 
It is well-known that there are two main classes of RSB, the one-step RSB (1RSB) and the full RSB (fRSB), and that these are characterised by a different structure of the metastable states, i.e.  marginal (and unstable under external perturbations) in fRSB, and well-shaped and surrounded by large barriers in 1RSB \cite{AspBraMoo04, CavGiaPar04, AnnGuaCav04}.

Many systems in different situations (physics, biology, ecology, neuroscience and economics) are intriguingly observed to lie just at the edge of stability or operate near criticality \cite{MullerWyart15,HerzHopfield95, Challet01, gene-Thurner08, BeggsTimme12, MoraBialek11,RMP,Moretti}. The recent application of RSB frameworks to some of such systems has been highly revealing, suggesting that they may dynamically select marginally stable states, as they could be the most numerous and/or those with the largest basins of attraction, underpinning a fRSB scenario 
\cite{BiroliCammarota18,Altieri-2021,Kaneko-2023,Pagnani-2006}.

Such a powerful framework, however, is only available for equilibrium systems, so it is limited to systems with reciprocal (or symmetric) interactions that preserve detailed balance. 
For systems with non-reciprocal there is, as yet, no general framework to quantify the rich diversity of emerging steady states, which can include limit cycles, chaotic trajectories  and fixed points ---although recent years have seen some progress in this direction \cite{FyoKhoPNAS16, ZamponiFolli19, FedFyoIps21, BertrandFyodorov22, ros-prl-2023}--- and there is still no method to predict the steady state that is {\it dynamically selected} by the system.

It is known that systems with asymmetric interactions can exhibit a transition to chaotic behaviour as the randomness in their interactions is increased \cite{Crisanti88,Sompo1} as well as exotic so-called {\emph{non-reciprocal}} phase transitions \cite{Non-reciprocal}, that constitute a topic of current research interest. 

In addition, it has been shown that, for asymmetric neural networks close to the transition into a chaotic phase, the mean number of fixed points diverges exponentially with network size, with the rate of divergence, called \emph{topological complexity}, following the trend of the largest Lyapunov exponent \cite{Wainrib-Touboul-13}. However, how topological complexity affects dynamical complexity remains unclear; indeed, a formal connection between spin-glass-like phases with exponentially many equilibria and chaotic dynamics has not been established yet. 

The aim of this manuscript is to contribute to the development of an interface between the statistical physics of classical disordered systems and dynamical systems theory, by showing that chaotic phases in non-equilibrium systems exhibit certain analogies with spin-glass phases in equilibrium (fRSB) systems.

We revisit a classical model for neural networks dynamics ---introduced  by Sompolinsky, Crisanti and Sommers in \cite{Crisanti88}--- and modify it to allow for interactions to be unbalanced (i.e. with a non-vanishing mean) and {\it randomly} asymmetric, as well as to incorporate "thermal noise" in the dynamics. 

We bypass the (hard) problem of calculating the topological complexity of different attractors in the configuration space 
 and focus on dynamical aspects (e.g. 
time-dependent moments) 
 of the trajectories at stationarity, 
 as originally proposed by SCS in \cite{Crisanti88}.

By computing two order parameters, namely the mean and variance of neural activity, we reveal the emergence of two types of criticality, one corresponding to ferromagnetic order and the other corresponding to spin-glass order. 
The transition from the disordered to the ordered phase can be rationalised in terms of spectral properties of the interactions matrix. 

Numerical analysis of the largest Lyapunov exponents (LLE) shows that chaotic dynamics arises in the spin-glass region of the phase diagram, for any value of the asymmetry of the couplings. 

For fully asymmetric couplings, we analyse the non-fixed point steady states and we determine the value of the variance that is {\it dynamically} selected by the system. 
We find that whenever the parameters allow for the existence of a multitude of physical (i.e. realizable) steady-state solutions, the system selects the steady state corresponding to the separatrix curve, 
which delimitates the basins of attractions of different steady states. This is reminiscent of the behaviour of fRSB systems at equilibrium,  
which select {\it marginally stable} states, i.e. saddles in the free-energy landscape, that are linked together by flat directions. Our analysis suggests that chaotic motion is the manifestation, at the level of {\it single network instances}, of such an ensemble averaged dynamics lying on the separatrix curve,  delimiting different {\it realizable} steady states. Such a motion is not time-translation invariant, similarly to ageing dynamics in spin-glass phases. 
Finally, we show that upon increasing the signal (i.e. \!imbalance of interactions) or the thermal noise the system is brought to a phase where only one bounded solution exists and chaos is suppressed. Such a behaviour can be seen again, as mirroring equilibrium phase transitions from spin-glass phases (with many equilibria) to ferromagnetic or paramagnetic phases (with a single equilibrium), when the signal or the noise, respectively, are raised above a critical value.

The manuscripts is structured as follows. In Sec. \ref{sec:model} we define the model. In Sec. \ref{sec:LSA} we study the linear stability of the quiescent state, which reveals the existence of two types of criticality. In Sec. \ref{sec:GFA} we analyse the dynamics of the model via a path integral formalism. In particular, we first consider the dynamics in the absence of noise, for uncorrelated interactions (Sec. \ref{sec:s0-g0}), then we turn to the noisy dynamics for uncorrelated interactions (Sec. \ref{sec:sn0-g0}). 
and finally we analyse the case of correlated interactions (Sec. \ref{sec:s0-gn0}). We discuss the main results and conclusions in Sec. \ref{sec:conclusions}. All technical details and complementary analyses are provided in the Appendices and the Supplementary Material (SM).

\begin{figure*}[tbh]
    \centering
    \includegraphics[width=1 \linewidth]{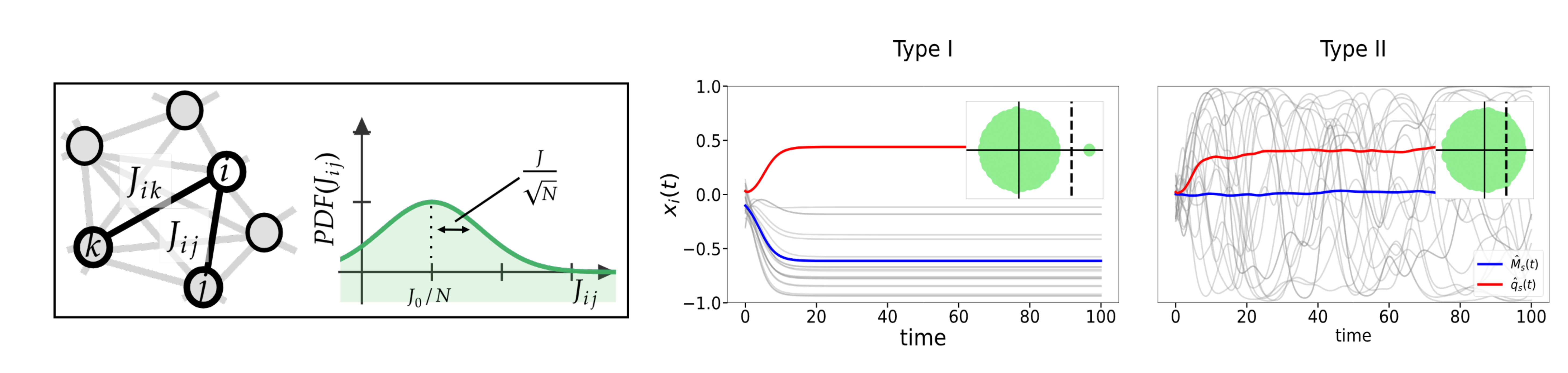}
    \caption{The left panel illustrates a fully-connected network with $N$ nodes and coupling strengths $J_{ij}$, drawn randomly and independently from a Gaussian distribution with mean $J_0/N$ and variance $J^2/N$.  The central and right panels sketch typical trajectories, $x_i(t)$, obtained from a numerical simulation of Eq. \eqref{eq:dynamics-def} for a network instance (with $\gamma=0$). The trajectories of a few nodes (solid grey lines) are plotted together with the mean $\hat{M}(t)=N^{-1}\sum_i x_i(t)$ (blue line) and variance $\hat{q}(t)=N^{-1}\sum_i x_i^2(t)$ (red line). For type-I phase transitions (central panel), the trajectories converge to steady-state values (on the same timescale) while for type-II (right panel) trajectories are highly irregular or chaotic. The insets show the distribution of eigenvalues (in the complex plane) for each case: for type-I transitions only one eigenvalue, the outlier, has crossed the instability line ${\rm Re}[\lambda]=1/g_c$ (shown by the dashed line), while for type-II, a section of the bulk has crossed the instability line (without a gap), giving rise to a more complex dynamics.}
    \label{fig:linear-analysis}
\end{figure*}

 \section{Rate model}
 \label{sec:model}
 
We consider a variant of the classical model introduced by Sompolinsky, Crisanti and Sommers (SCS) in \cite{Crisanti88}. The model consists of a fully-connected network with $N$ neurons (as shown Fig. \ref{fig:linear-analysis}, left panel), each one described by its time-dependent firing rate, $x_i(t)$, with $i=1,2,..., N$. Their dynamics obey the following set of coupled stochastic differential equations
\begin{equation}
    \dot x_i =-x_i +f \Big(g\sum_j J_{ij} x_j +\theta_i\Big)+\xi_i,
    \quad i=1,\ldots, N
    \label{eq:dynamics-def}
\end{equation}
where the first term on the right hand side describes a spontaneous decay of activity; the second one is the response to inputs, mediated by the gain function $f(x)$ which is a monotonically increasing function which saturates for large (in absolute value) arguments, taken here, without loss of generality, to be $f(x)=\tanh(x)$; $g$ is an overall coupling strength and the "synaptic weights", $J_{ij}$, are (quenched) Gaussian random  variables with mean $J_0/N$ and variance $J^2/N$
\begin{equation}
\overline{J_{ij}}=\frac{J_0}{N};~~~~\quad\overline{J_{ij}^2}-\frac{J_0^2}{N^2}=\frac{J^2}{N}
\label{weights}
\end{equation}
where overbars stand for network average.
We allow for correlations between $J_{ij}$ and its reciprocal, $J_{ji}$, 
\be
\overline{J_{ij}J_{ji}}-\frac{J_0^2}{N^2}=\gamma\frac{J^2}{N}
\ee
quantified by the Pearson's correlation coefficient $\gamma\in[-1,1]$, 
so that for $\gamma=1$, interactions are symmetric/reciprocal i.e. $J_{ij}=J_{ji}$, for $\gamma=0$ they are or uncorrelated or fully asymmetric/non-reciprocal,  and for $\gamma=-1$ they are anti-symmetric. The external field $\theta_i$ is added to generate response functions and will be set to zero afterwards. Finally, the third term, $\xi_i(t)$, is a zero-mean Gaussian white noise with $\bra \xi_i(t)\xi_j(t')\ket=2 \sigma^2 \delta_{ij}\delta(t-t')$, so that $\sigma$ is the  noise amplitude, and $\langle ... \rangle$ stands for noise average. 

Let us remark that this model is very similar to the classical one of SCS but differs in a few aspects: (i) the non-linear function applies to the sum of inputs rather than to each of them; this brings the model closer to rate models in neuroscience \cite{Dayan}; (ii) the connectivity matrix is allowed here to have a non-vanishing mean, so that one can study ``excitation" ($J_0>0$) and ``inhibition" ($J_0<0$) dominated regimes (we refer to these as ``unbalanced" cases); (iii)
interactions have an {\it arbitrary} degree of symmetry/reciprocity (encoded in $\gamma$); (iv) noise $\sigma^2\neq 0$ is included in the dynamics.
 In what follows, we discuss two alternative approaches to tackle analytically different aspects of the model: (i) a linear stability analysis allowing us to compute phase boundaries and (ii) a more complete path-integral formalism, allowing for a (dynamic mean-field) solution (exact in the infinite network-size limit).

\section{Linear-stability analysis}
\label{sec:LSA}

To scrutinize the possible dynamical phases of the system defined by Eq.(\ref{eq:dynamics-def}) we start by considering a linear stability analysis of the "quiescent" solution, where  $x_i(t) = 0, \, \forall i = 1, ..., N$ and the neural network remains inactive. For this we fix $\theta_i=0$,  $\sigma_i=0$, and expand $f(x)$ to the first order
\begin{equation*}
    \dot x_i =-x_i +g \,\sum_j J_{ij} x_j,
    \quad i=1,\ldots, N.
    \label{eq:dynamics-linear}
\end{equation*}
 Thus, the stability of the quiescent solution is controlled by the largest eigenvalue, $\lambda_M$, of the connectivity matrix $\mathbf{J}$, which depends on the model parameters $(J_0, J)$ as well as the degree of correlation, $\gamma$. In particular, the quiescent solution is stable if $ g \lambda_M<1$, and becomes unstable for values of $g$ above $g_c=1/\lambda_M$.

 The maximum eigenvalue can be easily computed (in the infinite network-size limit) using well-known results from random matrix theory (RMT) for the case of fully asymmetric ($\gamma=0$) and symmetric ($\gamma=1$) networks.
 In particular, for fully asymmetric networks with $\vert J_0 \vert< J$ the eigenvalues obey the Girko's circle law, meaning they 
 lay uniformly inside a circle of (spectral) radius $J$ in the complex plane centered at the origin \cite{vM,Doro, Tao}, i.e. the maximum eigenvalue coincides with $J$. If, however,  $\vert J_0\vert > J$, then one of the eigenvalues leaves the circle becoming a real-valued ``outlier" with value $J_0$ \cite{Tao,Rajan}. 
On the other hand, for symmetric networks, with $\vert J_0 \vert < J$, the eigenvalues are all real and obey the Wigner's semicircle law, so that they are uniformly distributed on a semicircle of radius $2J$ \cite{knowles_isotropic_2013}, i.e. the maximum eigenvalue coincides with $2J$. For symmetric networks with $\vert J_0 \vert > J$, the distribution of the bulk keeps following Wigner's law, but a single outlier eigenvalue appears at $J_0+J^2/J_0$ \cite{knowles_outliers_2014}. 

 Hence, the limit of stability, $g_c$, is given by the largest element between the outlier and the largest eigenvalue in the bulk, i.e. one has, for $\gamma=0$
\begin{equation}
  g_c ~{\rm max} (J,J_0) =1 \rightarrow  \frac{1}{g_c J}={\rm max}\Big(1, \frac{J_0}{J}\Big)
\label{eq:bifurcation_Linear_0}
\end{equation}
while, for $\gamma=1$ 
\begin{equation}
 g_c ~{\rm max} \left(2J,J_0+\frac{J^2}{J_0}\right)=1 \rightarrow   \frac{1}{g_c J}={\rm max}\Big(2, \frac{J_0}{J}+\frac{J}{J_0}\Big).
\label{eq:bifurcation_Linear_1}
\end{equation}

Furthermore, we note that for general $\gamma$'s, the distribution of eigenvalues has also been derived for balanced matrices in \cite{sommers_spectrum_1988}, while for unbalanced matrices the position of the outlier has been calculated in \cite{orourke_low_2014}.

Thus, depending on whether the largest eigenvalue is an outlier or is at the edge of the bulk, two different paths to instability emerge.
If the outlier is the largest eigenvalue, as $g$ is increased, a single collective mode (given by the eigenvector associated to the outlier eigenvalue) destabilizes, as illustrated in the central panel of Fig.\ref{fig:linear-analysis}. In this case, the system
reaches a fixed steady state above the instability threshold. If, on the other hand, the maximum eigenvalue lies at the edge of the bulk, a continuum of modes may become destabilized as $g$ is increased across the edge of instability, as illustrated in Fig.\ref{fig:linear-analysis}, right panel. In this case, the dynamics of the system turns out to be non-trivial and complex trajectories emerge.
This difference is the core of the distinction between type-I and type-II criticalities \cite{Helias2019, Galla-2023}. 
  
Let us emphasize that the linear stability analysis does not indicate how to characterize the two types of criticality e.g. in terms of order parameters and critical exponents, nor how to locate the transition between type-I and type-II phases.
Thus, to tackle analytically this general problem one needs to resort to a more sophisticated dynamical mean-field approach as developed in the following section.

\section{Path-integral formalism}
\label{sec:GFA}

To analyze the full phase diagram one can resort to a ``dynamic mean-field" approach (DMF) as originally introduced by SCS \cite{Crisanti88} and later developed in a number of works (e.g, \cite{Helias2018,Sompo-PRX15, rajan_stimulus-dependent_2010,F+Ostojic}).
Alternatively, one can employ a more systematic path-integral formalism (or generating functional analysis) as originally introduced by Martin, Siggia and Rose \cite{MartinSiggiaRose}
and successfully applied to neural networks in \cite{Coolen2001, HatchettCoolen2004, Crisanti18}, which allows to formally derive the DMF equations and quantify their limits of validity \cite{Helias2018,Crisanti18}. The calculation is quite standard but, for the sake of completeness, here we reproduce the main steps.

The starting point of the method is the definition of the moment-generating functional of the random (vector) function $\bx(t)$
\be
Z[\bpsi]=\left \langle e^{i~\sum_i\int dt\, x_i(t)\psi_i(t)} \right \rangle
\label{eq:Z-def}
\ee
where 
the average $\bra \ldots \ket$ is taken over the probability distribution $P[\bx(t)]$ of 
trajectories $\bx(t)$, generated by Eq.(\ref{eq:dynamics-def}), for a fixed (quenched) connectivity matrix $\bJ$. From such a generating functional one can easily compute the chief quantities describing the system's collective dynamics, such as the mean activity of a neuron $i$
\be
M_i(t)=\bra x_i(t)\ket=\left. \frac{\delta Z[\bpsi]}{i\delta \psi_i(t)}\right|_{\bpsi={\bf 0}},
\label{eq:m_gfa}
\ee
the two-time pairwise correlations 
\begin{equation}
    C_{ij}(t,t')=\bra x_i(t) x_j(t')\ket=
    \left. \frac{\delta Z[\bpsi]}{i\delta \psi_i(t)i\delta\psi_j(t')}\right|_{\bpsi={\bf 0}},
    \label{eq:C_gfa}
\end{equation}
and the response functions
\begin{equation}
R_{ij}(t,t')=\frac{\delta \bra x_i(t)\ket}{\delta \theta_j(t')}=\left. \frac{\delta^2 Z[\bpsi]}{i\delta\theta_j(t')\delta \psi_i(t)}\right|_{\bpsi={\bf 0}}.
\label{eq:R_gfa}
\end{equation}

In order to compute disorder-averaged quantities, one needs to average the generating functional $Z[\bpsi]$ over the distribution $P(\bJ)$. This procedure leads ---once the infinite network-size ($N\to\infty$) limit has been taken--- to a self-consistent stochastic equation for the  dynamics of a {\it representative} neuron, called dynamical-mean-field equation \cite{Crisanti88,Helias2018,Helias-book}, see Supplemental Material (SM) Sec. \ref{app:GFA} for further details:
\begin{widetext}
\begin{equation}
\dot x(t)=-x(t)+ \tanh \Big[ J_0g M(t)  + \gamma J^2 g^2 \int dt'\, R(t,t')x(t') + 
\theta(t) +\phi(t) \Big]+\xi(t).
\label{eq:single-neuron} 
\end{equation}
\end{widetext}
In this equation, $\xi(t)$ is a zero-mean Gaussian white noise with $\bra \xi(t)\xi(t')\ket=2\sigma^2 \delta(t-t')$, $\phi(t)$ is a random Gaussian field with zero mean and auto-correlation 
\begin{equation} \label{eq:xi-phi}
\langle \phi(t) \phi(t')\rangle = J^2 g^2 C(t, t')
\end{equation}
and the order parameters $M(t)$, $C(t,t')$ and $R(t,t')$ need to be calculated self-consistently, as averages over realizations of the effective single-neuron process:
\begin{eqnarray}
&& M(t) =\bra x(t)\ket,
\\
&& C(t,t')=\bra x(t)x(t')\ket
\\
&& R(t,t') = \frac{\delta \bra x(t)\ket}{\delta \theta(t')}
= \bra \frac{\delta x(t)}{\delta \phi(t')}\ket,
\label{eq:Delta_C}
\end{eqnarray}
where in the second equality of the last equation we have used that 
$\theta$ and $\phi$ have the same role in Eq.\eqref{eq:single-neuron}.  
The random function $\phi(t)$ can be interpreted as an  ``interference" term that quantifies the impact, on the dynamics of the representative neuron, of all the other neurons that interact with it, while the second term in the square brackets of Eq.\eqref{eq:single-neuron} accounts for the so-called ``retarded" self-interactions of the representative neuron with its own past, i.e. past values $x(t')$ influence $x(t)$ at later times $t>t'$. This is due to the fact that the representative neuron sends signal to its neighbours through links which are {\it correlated} with those used by the neighbours to send signal back to it. We note that (as usual) this term only arises when such correlations in the links are present (i.e. $\gamma\neq 0$) and it makes the effective dynamics non-Markovian. From here on, the external field is fixed to zero, $\theta(t)=0$.

In the following sections, we investigate separately the cases of (i) uncorrelated interactions ($\gamma=0$) with 
noiseless dynamics ($\sigma=0$); (ii) 
uncorrelated interactions ($\gamma= 0$) with 
noisy dynamics ($\sigma\neq 0$) and (iii)
 correlated interactions ($\gamma\neq 0$), focusing for simplicity on the noiseless case.


\begin{figure*}[tbh]
    \centering
    \includegraphics[width=1 \linewidth]{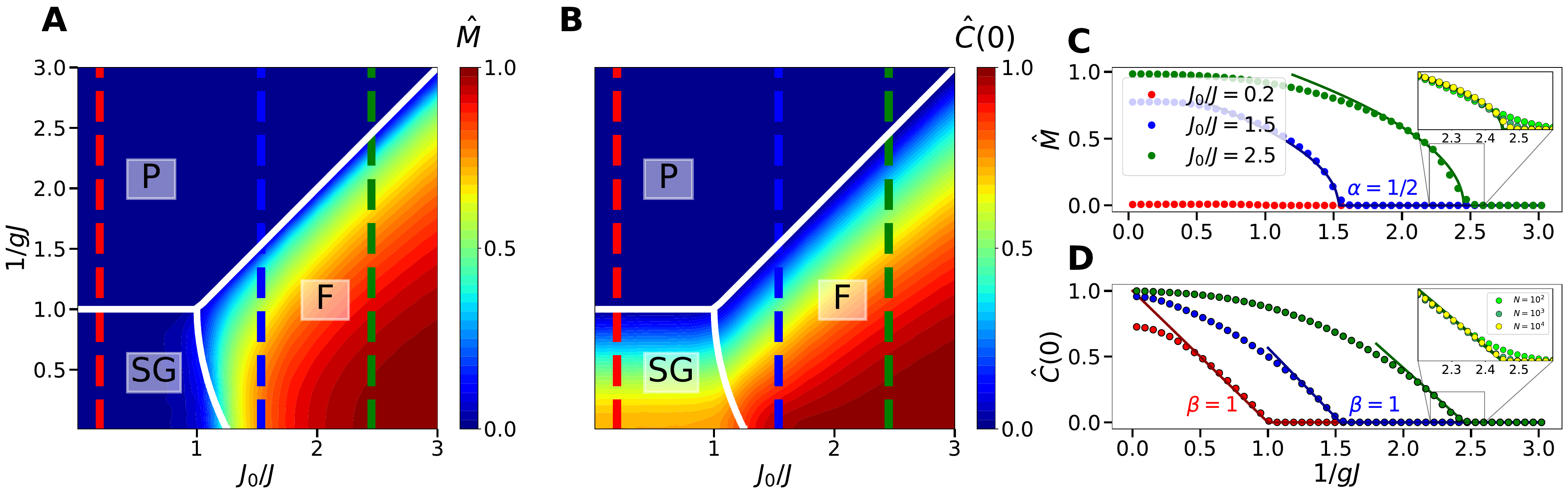}
    \caption{Phase diagram of the model for \textbf{uncorrelated} ($\gamma = 0$) and \textbf{noiseless} dynamics ($\sigma^2 = 0$). 
    \textbf{Panels (A, B)}: Heat map of the time-averaged mean activity $\hat M$ (panel (A)) and mean-squared activity (or equal-time correlator) $\hat C(0)$ (panel (B)), obtained from simulations as a function of the control parameters $J_0/J$ (as color coded) and $1 / g J$.
    The white lines represent the theoretically predicted critical curves (from stability analyses of fixed-point solutions) separating the paramagnetic (P), ferromagnetic (F) and spin-glass (SG) phases. \textbf{Panels (C, D)}: Symbols show $\hat M$, from panel (A), and $\hat C(0)$, from panel (B), obtained from simulations, versus $1/gJ$, at three different values of $J_0/J$, corresponding to the three dashed vertical lines on (A) and (B). Solid lines show the theoretically predicted asymptotic behaviour around the critical point such that $M \sim 
    (g-g_c)^{\alpha}$ (defined only for the ferromagnetic transition) and $q \sim 
    (g-g_c)^{\beta}$. (Note that $M$ and $q$ denote, respectively, the values of the mean and the 
    mean-squared activity obtained from the theory ---by averaging over initial conditions and network ensemble---, assuming a fixed-point solution, whereas $\hat M$ and $\hat C(0)$ denote their numerical counterparts obtained from numerical simulations by averaging over different realizations.)
    The inset illustrates the effect of increasing the system size ($N = 100, 1000, 10000$). }
    \label{fig:phase-diag-asym}
\end{figure*}

\subsection{Uncorrelated interactions ($\gamma=0$) and noiseless dynamics ($\sigma=0$)}
\label{sec:s0-g0}

For uncorrelated or ``fully asymmetric'' couplings (i.e. $\gamma=0$), the equation of motion for the noiseless dynamics ($\sigma=0$) takes a particularly simple form
\bea
\dot x(t)=-x(t)+\tanh [J_0g M(t) +\phi(t)],
\label{eq:ODE_x_mphi}
\eea
that can be  explicitly analyzed.

\subsubsection{Fixed-point solutions}
\label{sec:FP-s0-g0}

Let us start by assuming that the system reaches a single stable fixed point $x(t)=x$. In such a case, $\bra x\ket=M$, $\bra x^2\ket=C(0)=q$, and each realization of $\phi(t)$ becomes 
a static zero-mean Gaussian random variable with variance $J^2 g^2 q$ 
\cite{OpperDiederich-PRL92, SidhomGalla-PRE20}. Therefore, imposing the stationarity condition, $\dot x=0$, in Eq.(\ref{eq:ODE_x_mphi}) and averaging over $\phi$ one readily obtains: 
\bea
M&=&\bra \tanh(J_0 g M + \phi)\ket 
\nonumber\\
&=&\int \D\psi
\tanh(J_0gM+gJ \sqrt{q}\psi)\,
\label{eq:M-self-asym}
\eea
where  the short-hand notation
$\D\psi=e^{-\psi^2/2}d\psi/\sqrt{2\pi}$ has been introduced. Similarly, the variance $q$ 
at the fixed-point solution  is
\bea
q&=&\bra \tanh^2(J_0 g M+\phi)\ket 
\nonumber\\
&=&\int \D\psi
\tanh^2(J_0 g M+g J \sqrt{q}\psi).
\label{eq:q-self-asym}
\eea
An alternative, simpler derivation of Eqs.(\ref{eq:M-self-asym})-(\ref{eq:q-self-asym}) ---not generalizable to the $\gamma \neq 0$ case--- is presented in SM (Sec. \ref{app:naive}).

Before proceeding, we note that Eq.(\ref{eq:M-self-asym}) and Eq.(\ref{eq:q-self-asym}) are identical to the equations for the magnetization and the Edward-Anderson order parameter, respectively, in the well-known Sherrington-Kirkpatrick (SK) model for spin-glasses \cite{sherrington_solvable_1975}, within a replica symmetric ansatz, with $1/g$ playing the role of the temperature (see e.g. Eqs.(2.28)-(2.30) in \cite{Nishimori}).
Such an equivalence, at the level of order parameters, between the SK model and the neural-network rate  model under study, is surprising as the two models are defined in rather different ways. 
In particular, the SK model is defined for Ising (i.e. $\pm 1$) variables with symmetric interactions while here Eq.\eqref{eq:M-self-asym} and Eq.\eqref{eq:q-self-asym} have been derived for continuous state variables and non-symmetric interaction matrices.

From the analogy with spin glasses, 
 Eq.\eqref{eq:M-self-asym} and Eq.\eqref{eq:q-self-asym} are well-known to have a trivial disordered  (or "paramagnetic" in the language of magnetic systems) solution with $(M,q)=(0,0)$, which is stable at sufficiently high temperature (here corresponding to small values of the  coupling constant $g$). From the disordered phase two different types of order may emerge, depending on the relative values of the mean $J_0$ and variance $J$ of the coupling distribution.

In particular, for $J_0>J$, there is a bifurcation at $g J_0 = 1$ from the disordered phase to an ordered (or ``ferromagnetic") phase in which the up-down symmetry is broken, i.e., there is a non-vanishing magnetization, $(M \neq 0, q \neq 0)$.
On the other hand, for $J_0<J$, a bifurcation occurs at $gJ=1$ from such a disordered phase to a ``spin-glass" phase $(M=0,q\neq0 )$
\cite{sherrington_solvable_1975}.
Therefore, the disordered phase loses its stability at the critical value $g_c$, given by Eq.\eqref{eq:bifurcation_Linear_0}, in agreement with results from linear-stability analysis.
In particular, type-I criticality discussed in Sec. \ref{sec:LSA} can be identified with the transition from the paramagnetic to the ferromagnetic phase, while type-II criticality corresponds to the transition from the paramagnetic to the spin-glass phase. Such criticalities can therefore be fully characterised by the order parameters $M$ and $q$, which 
correspond to mean neural activity and mean-squared activity, respectively.
In addition, the critical line separating the ferromagnetic from the spin-glass phase 
can be obtained by analyzing the limit of linear stability of the $M=0$ solution, which ---as derived in the SM (Sec.  \ref{app:FP_diagram_gamma0})--- leads to
\be
\frac{1}{gJ}=\frac{J_0}{J}(1-\qt^*)
\label{eq:SG-FM}
\ee
where $\qt^*$ is the solution of 
\be \label{eq:q-sg-fm}
\qt^* = \int \D\psi \tanh^2\Big(g J \sqrt{\qt^*}\psi\Big).
\ee

The resulting phase diagram, in the parameter space ($J_0/J ,  1/g J$), is shown in 
Fig.\ref{fig:phase-diag-asym}, where the solid white lines in panels (A) and (B) represent the critical lines separating the paramagnetic (P), ferromagnetic (F) and spin-glass (SG) phases.

Expanding the self-consistency Eqs.\eqref{eq:M-self-asym}-\eqref{eq:q-self-asym} to the first sub-leading order, 
it is possible to obtain the critical exponents $\alpha$ and $\beta$ such that $M \sim 
(g-g_c)^{ \alpha}$ and  $q \sim 
(g-g_c)^{ \beta}$ (see SM, Sec. \ref{app:FP_diagram_gamma0}). For the ferromagnetic transition $J_0 > J$, one has 
$\alpha = 1/2$ and $\beta=1$ (as in the Ising model) whereas for the SG transition, i.e. for $J_0 < J$,
the exponent $\alpha$ is trivially $0$ and $\beta = 1$.

In addition to fixed-point solutions, a rich set of possible non-fixed-point solutions is expected to appear with non-symmetric interactions, including limit cycles and chaotic
trajectories \cite{Crisanti88,Sompo-PRX15,Crisanti18}.
Among these, stationary solutions are characterised by a time-independent average $M(t)=M$ and a time-translation invariant two-time correlator $C(t,t')=C(t-t')$, which implies a time-independent equal-time correlator $C(t,t)=C(0)$ (note that for fixed-point solutions $C(0)=q$, however for non-fixed-point solutions $C(0)\neq q$). Whenever the steady state is not a fixed point, node activities $x_i(t)$ fluctuate in time, even at stationarity. One can therefore ask whether the phase diagram derived for fixed-points gives reliable information on the phase transition behaviour of the system, whose dynamics is not necessarily at a fixed point. 

To answer this question, we ran computer simulations of the microscopic dynamics and ---assuming that the system eventually reaches a steady-state--- we computed the stationary magnetization and equal-time (or zero-lag) correlator as the sample average, for $S$ 
independent simulated trajectories ${\bf x}_s(t)$, with $s=1,\ldots, S$, of the mean activity \footnote{We use the ``hat'' notation for quantities measured in simulations, to distinguish them from their theoretical counterparts, as obtained from 
the path integral approach. In particular, single-realizations are denoted by a subscript $s$, e.g., $\hat{M}_s$, whereas averages over many realizations are denoted with $\hat M$.}

\be
\hat M_s=\frac{1}{t_m}\sum_{t=t_0}^{t_0+t_m}\frac{1}{N}\sum_{i=1}^N x_{i,s}(t)
\ee
and mean-squared activity
\be
\hat C_s(0)=\frac{1}{t_m}\sum_{t=t_0}^{t_0+t_m}\frac{1}{N}\sum_{i=1}^N x^2_{i,s}(t)
\label{eq:Cs}
\ee
respectively, where
$t_m\gg 1$ and $t_0$ is a sufficiently long time to allow the dynamics to relax to stationarity.
 Fig.\ref{fig:phase-diag-asym} shows a heat map of their averages 
$\hat M=S^{-1}\sum_{s=1}^S \vert\hat M_s \vert$ (panel (A)) and
$\hat C(0)=S^{-1}\sum_{s=1}^S \hat C_s(0)$ (panel (B)), where the absolute value $\vert ... \vert$ is used to avoid the cancellation between positive and negative states of the system. 
Simulations are performed with $S=1000$ realizations, for a system with size $N=1000$,   
for different values of the parameters $(J_0/J,1/gJ)$. 
Notably, the phase diagram derived for fixed-point solutions captures very well the phase transition behaviour of $\hat C(0)$ and $\hat M$, which vanish in the disordered
(paramagnetic) phase and become non-zero in the ordered (spin-glass and ferromagnetic) phases. In addition, their asymptotic behaviour close to criticality matches the scaling theoretically predicted for fixed-point solutions (see Fig.\ref{fig:phase-diag-asym} panels (C) and (D) and SM, Sec.~\ref{app:FP_diagram_gamma0} for details). This suggests that close to the instability line of the paramagnetic phase, trajectories are at a fixed point. This is consistent with bifurcation analysis, showing that the first non-trivial 
steady-state solutions to appear below criticality are fixed-point solutions (see SM, Sec. \ref{app:SS_diagram_gamma0} for details).  
In the next subsection, we study in detail non-fixed-point solutions at stationarity.

\subsubsection{Non-fixed point steady states}
\label{sec:NFP-s0-g0}

Let us now focus on stationary solutions other than fixed points. These are characterised by a constant first moment $M(t)=M$ and time-translation invariant correlator $C(t,s)=C(t-s)$, where $M$ is given by 
\bea
M&=&\int \D\psi
\tanh(J_0gM+gJ \sqrt{C(0)}\psi)
\label{eq:M-self-asym-SS}
\eea
with $C(0)$ denoting the equal-time correlator, while the two-time correlator $C(\tau)$ satisfies  
\bea
(1-\partial_\tau^2) \, C(\tau)&=&\Xi(C(\tau),C(0),M)
\label{eq:dot_C}
\eea
with
\begin{equation}
\Xi(C,C(0),M) = \int \mathcal{D}_{\phi\phi'} \tanh[J_0 g M+\phi]\tanh[J_0 g M+\phi'] 
\label{eq:Xi}  
\end{equation}
where
\be \label{eq:Gaussian_kernel}
\mathcal{D}_{\phi \phi'} = \frac{d\phi d\phi'}{2\pi g^2 J^2\sqrt{\det \bC(\tau)}} \exp \left(-\frac{\bphi^T \bC^{-1}(\tau)\bphi}{2 J^2 g^2} \right), 
\ee
see SM, Sec. \ref{app:C-dynamics-SS} for details. 

\begin{figure*}
\includegraphics[width=1 \linewidth]{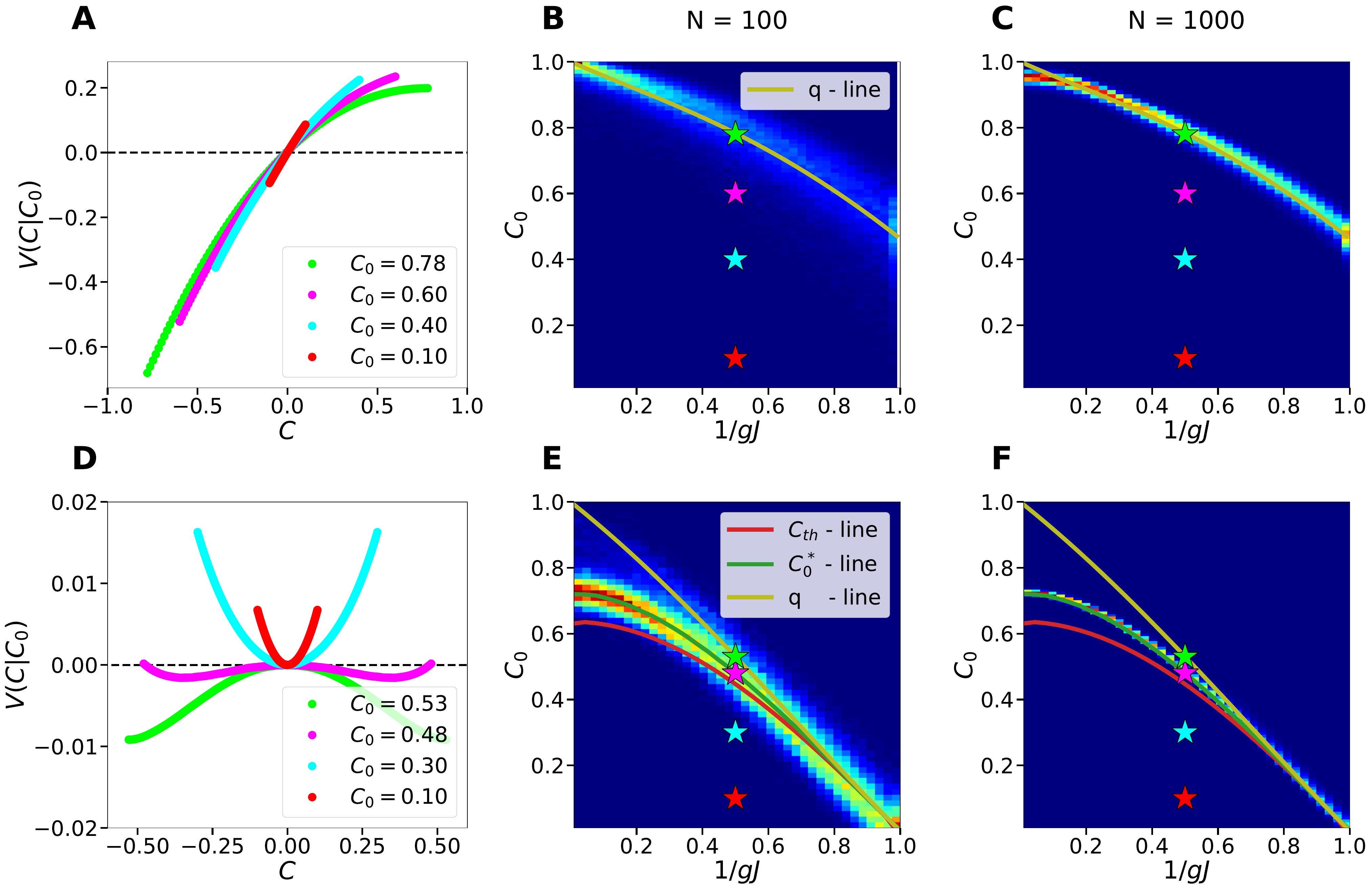}
    \caption{Analysis of the motion in the effective potential $V(C|C(0),M)$ 
    for the \textbf{uncorrelated} ($\gamma = 0$) and \textbf{noiseless} ($\sigma^2 = 0$) case. {\bf Panels (A, D)}: potential $V(C|C(0),M)$ 
    as a function of $C$ for $M\!\neq\! 0$ (ferromagnetic phase, with $J_0/J = 1.5$) and $M\!=\!0$ (spin-glass phase, with $J_0/J = 0.5$), respectively. Each curve is obtained for a different value of $C(0)$ (as shown in the legend with a color code) and is plotted in the range $C\in[-C(0),C(0)]$. Note that only in the SG phase the potential may exhibit a shape with either one or two wells.
    The heat maps in panels {\bf (B, C, E, F)}
    describe the probability distribution function (PDF) $P(C_0)$ 
     (as defined in Eq.\eqref{eq:PDF_Cs})
     of finding a given value of $C(0)$ for the ferromagnetic case \textbf{(B, C)} and the SG phase \textbf{(E, F)}, for different values of $1/gJ$ (obtained from $S=1000$ simulations of the microscopic dynamics, averaged over a time window $t_m=2000$, for system size $N=100$ (B, E) and $N=1000$ (C, F)). Each simulation corresponds to a different realization of the random initial condition and quenched disorder. Stars show the values of $1/gJ$ and $C(0)$ used to plot panels (A, D). 
    In panels (B, C), the solid line ("q-line") describes the solution $q$ obtained from the fixed-point solution, revealing that for large $N$, the PDF becomes peaked around this value, so that $C(0)=q$. On the other hand,  in panels (E, F), the PDF becomes more and more peaked (as the system size is enlarged from $N=100$ to $N=1000$) around a value $C_0^\star \neq q$, corresponding to the separatrix point (lying in the green curve) at which the potential verifies $V(C(0) | C(0), 0) = 0$ ($C(0) = 0.48$ in (D)). Note that such a green line is in between the q-line (yellow) and the threshold curve $C_{\rm th}$ (red)
    as described in the main text, so that in particular $C(0) \neq q$ in the SG phase.}
 \label{fig:non_fixed_points}
\end{figure*}

For ${\bf C}$ to be positive definite, we must have $C^2(0)\geq C^2(\tau)$, i.e. $C(0)\geq |C(\tau)|$, as expected for a stationary process, where the correlation function (at any time-lag $\tau$ larger than $0$) is bounded to be smaller than the variance. An expression for the variance $C(0)$ is provided in the SM (Sec. \ref{app:C-dynamics-SS}), however, let us remark that it requires the knowledge of the time-dependent (or ``non-persistent") order parameter $C(\tau)$, consistently with earlier findings in asymmetrically diluted recurrent neural networks 
\cite{HatchettCoolen2004}. Hence, the equations for the time-invariant (or ``persistent") order parameters $M$ and $C(0)$ are not a closed set (not even at stationarity),
when the system is not at a fixed point. Only when the system is at a fixed point, i.e. $x(t)=x$, then $C(0)=q$ and the equations for $M$ and $q$ form a closed set.

Following \cite{Crisanti18}, we note that Eq.\eqref{eq:dot_C} can be cast in a gradient-descent equation
\be
\partial_\tau^2 C(\tau)=-\frac{\partial V(C|C(0),M)}{\partial C}
\label{eq:motion_in_V}
\ee
on the potential 
\be
V(C|C(0),M)=-\frac{C^2}{2}+
\int_0^{C} dC'\,\Xi(C',C(0),M)
\ee
which depends on the persistent order parameters  $M$ and $C(0)$.

Note that for  $J_0=0$, the previous potential reduces to the one governing the correlation function in the model of SCS \cite{Crisanti88}. Although ---given that, unlike ours, in their model the sum over $j$ in Eq.\eqref{eq:dynamics-def} appears outside the hyperbolic tangent--- such an equivalence might seem surprising  it can be rationalised by a simple argument as explained in App.\ref{app:equivalence}. 

As already noticed in \cite{Crisanti18}, solutions of Eq.\eqref{eq:motion_in_V} conserve the total energy 
\be
E=\frac{1}{2}\dot C^2+V(C|C(0),M).
\label{energy}
\ee
Moreover, at stationarity, $C(\tau)=C(-\tau)$, so that $\dot C(0)=0$ and, hence, the initial kinetic energy vanishes. Thus, from energy conservation one has
\be
\frac{1}{2}\dot C^2(\tau)+V(C|C(0),M)=V(C(0)|C(0),M).
\ee

Physical solutions must be bounded, i.e. $|C|\leq C(0)$, this requires $V'(C(0)|C(0),M)>0$ and $V'(-C(0)|C(0),M)<0$ at the two boundaries $C=\pm C(0)$, respectively (since $\dot C(0)=0$).
It can be easily shown that, for any $M$, 
$V'(C(0)|C(0),M)<0$ if $C(0)>q$, hence such initial conditions are unphysical: 
the system will have to select a stationary value of the equal-time correlator that is smaller or equal than the persistent parameter $q$.

Let us now study separately the cases (a) $M\neq0 $ and (b) $M=0$:

\paragraph{\textbf{For $\mathbf{M \neq 0}$.}}  
One can show (see SM \ref{app:NP-g0-s0} C) that the potential is a monotonic non-decreasing function for any initial condition $C(0)\leq q$ and that there is only one stationary point, at the boundary of the physical region $C=C(0)=q$ (see Fig.\ref{fig:non_fixed_points}, panels (A)). Therefore, the only possible bounded steady-state solutions for $M\neq 0$ are fixed-point solutions with $C=q$. 
 
To verify this, we ran $S$ simulations of the microscopic dynamics to obtain trajectories ${\bf x}_s(t)$, with $s=1\ldots S$ and computed, for each realization $s$, the equal-time correlator at stationarity $\hat{C}_s(0)$, as defined in Eq.\eqref{eq:Cs}. The resulting distribution
\be
P(C_0)=\frac{1}{S}\sum_{s=1}^S\delta(C_0-\hat{C}_s(0))
\label{eq:PDF_Cs}
\ee
is plotted in  Fig.\ref{fig:non_fixed_points}(B, C) as a heat map, for different values of $1/gJ$, two different network sizes $N$, and a fixed value of $J_0/J$ (corresponding to the  ferromagnetic region, i.e. $M \neq 0$). Results show that, as the network size $N$ is increased, the probability density concentrates on the curve corresponding to $C(0)=q$ (solid line), confirming that, indeed,  the value of $C(0)$ selected dynamically by the system coincides with $q$.

\paragraph{\textbf{For $\mathbf{M = 0}$.}}
One can show that ---as illustrated in Fig.\ref{fig:non_fixed_points} panel (D)--- $V$ has the shape of a double-well potential for any initial condition $C(0)$ smaller than $q$ but larger than a certain threshold value $C_{\rm th}$ solution of
\bea
\int \D\psi \tanh^2(gJ\sqrt{C_{\rm th}}\,\psi)=1-\frac{1}{gJ},
\label{eq:cth}
\eea
 whereas below such a threshold ($C(0)<C_{\rm th}$) it becomes a single-well potential (see SM \ref{app:NP-g0-s0} C for further details).  
 
 Therefore, any initial condition $0\leq C(0)\leq q$  leads to bounded solutions, but ---depending on the value of $C(0)$ that is dynamically selected by the system--- different steady-state solutions may arise as summarized in the following:

\begin{itemize}
\item For $0<C(0)<C_{\rm th}$ the potential has a single-well shape, so that the motion is periodic between $C(0)$ and $-C(0)$. 

\item For $C_{\rm th}<C(0)<q$, the potential has a double-well shape with a local maximum at $C=0$, and three possible types of dynamics can emerge: 

\begin{itemize}
\item(i) periodic motion in one of the single wells corresponding to motion below the separatrix curve ($C(0)> C_0^*$ such that $V(C(0)|C(0),0)<0$); 

\item (ii) asymptotic motion towards $C=0$, corresponding to motion on the separatrix curve ($C(0)=C_0^\star$ such that  $V(C_0^*|C_0^*,0)=0$); 

\item (iii) periodic motion between $C(0)$ and $-C(0)$, i.e. covering the two wells, corresponding to motion above the separatrix curve ($C(0) < C_0^*$ such that $V(C(0)|C(0),0)>0$).    
\end{itemize}

\item For $C(0)=q$, the system remains at the fixed point  $C(\tau)=q$ for all values of $\tau$, i.e. the solution is a fixed point.
\end{itemize}

These different conditions are illustrated in Fig.\ref{fig:non_fixed_points} (E, F), which shows $C_{\rm th}$, $C_0^\star$ and $q$ as a function of $1/gJ$ and $C(0)$ (red, green and yellow solid lines, respectively), at a fixed value of $J_0/J$ (corresponding to the spin-glass region, $M=0$) and for two different system sizes (panel (E), $N=100$; panel (F), $N=1000$). 

The question that remains to be answered is: what is the value 
of the equal-time correlation, $C(0)$, that is selected by the system 
at stationarity? 
To ascertain this, we plot, in the same figure, the probability distribution  function Eq.\eqref{eq:PDF_Cs}, computed from $S=1000$ simulations, versus $1/gJ$, as a heat map, for two different 
system sizes. These results clearly reveal that, as the system size $N$ increases, 
the probability density concentrates on the curve $C(0)=C^*_0$. 
This means that in the thermodynamic limit  the time-average $C_s(0)$ of any single instance $s$ of the system (obtained for a given initial condition and a particular realization of the quenched disorder) settles precisely on the separatrix curve, i.e. at the boundary that delimits the basins of attraction of the two potential minima $\pm \bar C(C_0^*)$. Therefore, the solution sits on an unstable stationary state, where any slight perturbation causes it to move away from the separatrix in either of the two adjacent states. 
\begin{figure}[h]
\includegraphics[width= \linewidth]{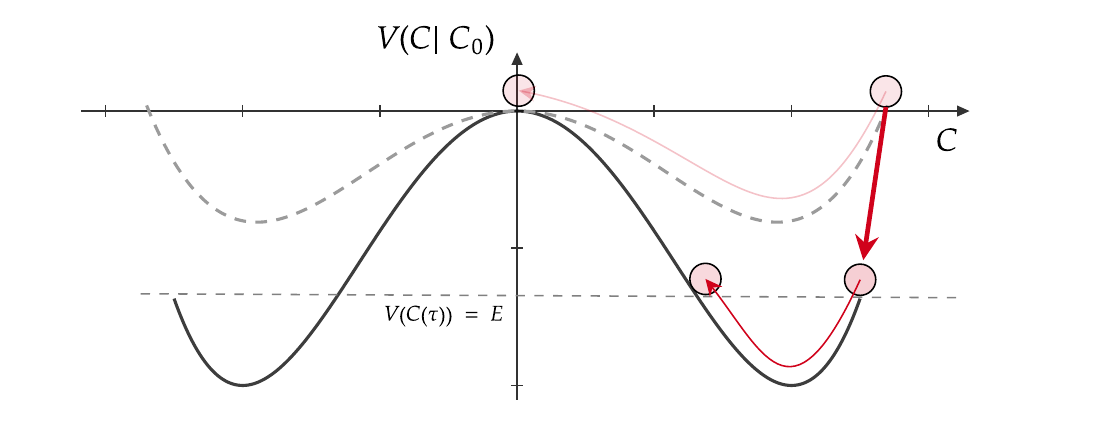}\\
\includegraphics[width=\linewidth]{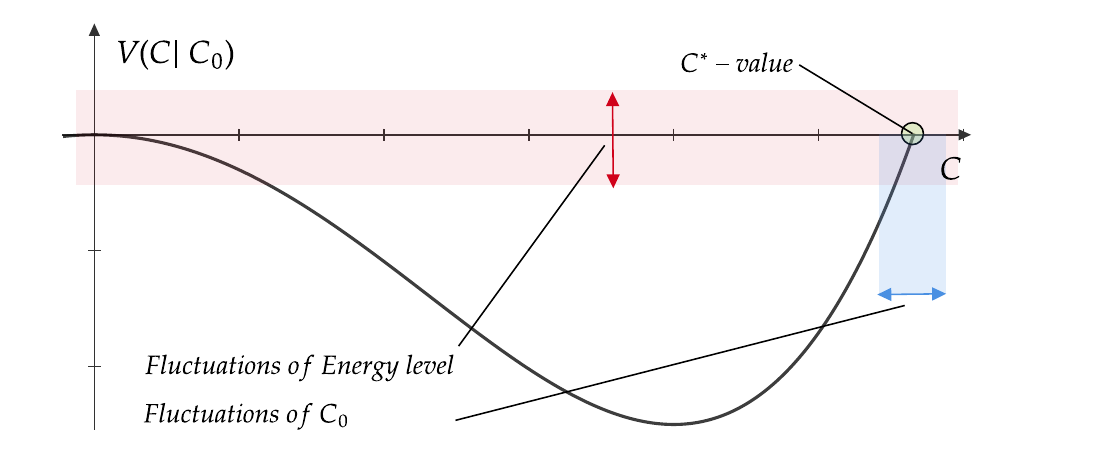}
\caption{Illustration of how fluctuations in the initial condition translate into changes in the shape of the potential (upper). Small fluctuations have a large impact when the initial condition varies around the separatrix value $C^*_0$ (lower).}
\label{fig:potential}
\end{figure}

\begin{figure*}[tbh] \large 
\includegraphics[width=01 \linewidth]{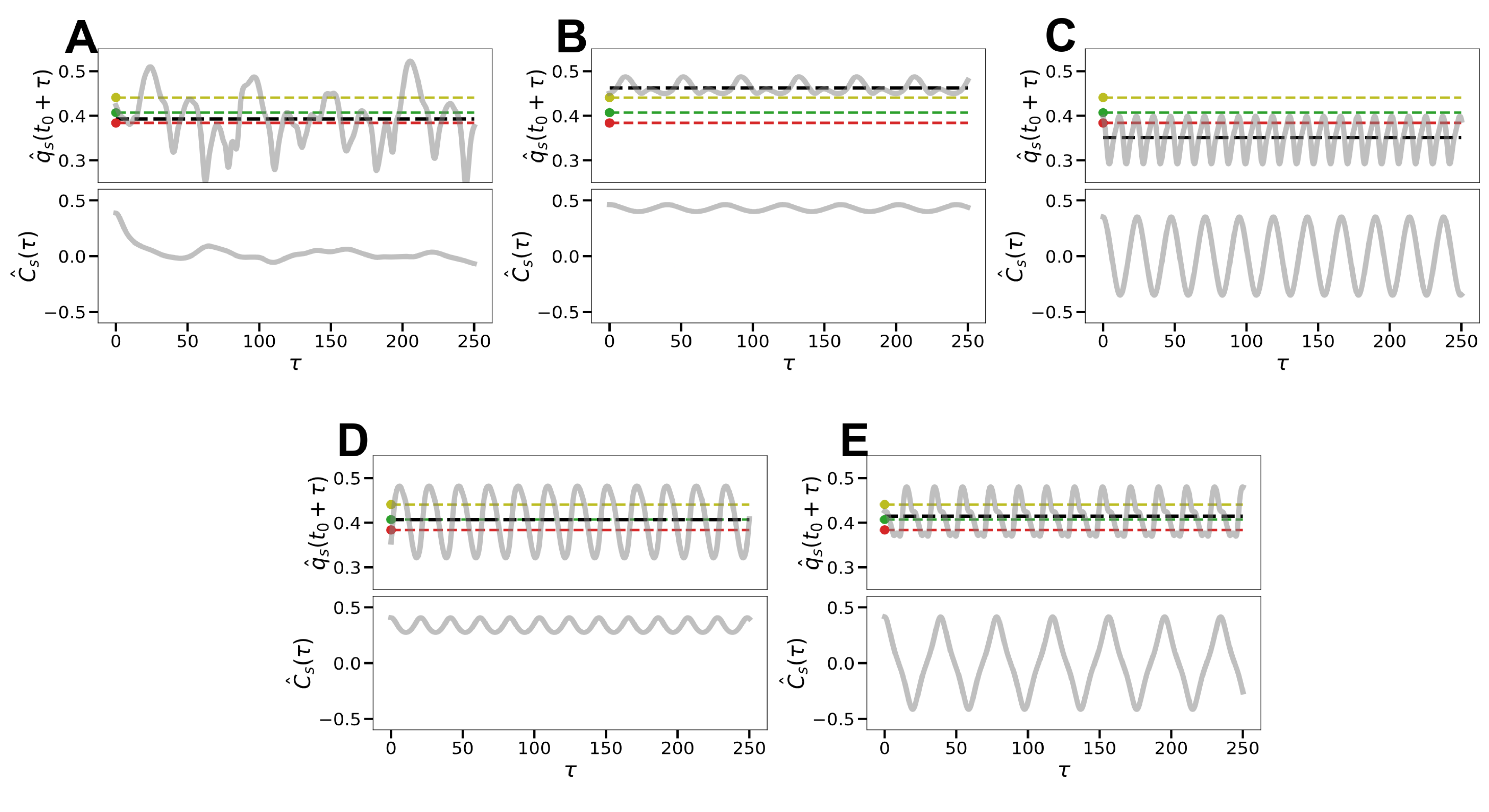} 
 \caption{Single instances of the dynamics for \textbf{uncorrelated} ($\gamma = 0$) and \textbf{noiseless} dynamics ($\sigma^2 = 0$). The  
 (upper panels) show the time-dependent mean-squared activity $q(t_0+\tau)=N^{-1}\sum_i x_i^2(t_0+\tau)$ (gray line in the upper panels) plotted together with its time average $\hat C_s(0)$ and (lower panels) $\hat C_s(\tau)$. Curves are obtained from simulations with system size $N=100$ and $1/gJ=0.58$ (i.e., within the SG phase). Observe that the trajectories reveal either chaotic behavior (A) or periodic motion (B, C, D, E). The yellow and the red dashed lines show $q$ and $C_{\rm th}$, respectively while the green dashed line --in between the previous two---
 marks the separatrix $C_0^\star$.
 In (A) the mean-squared activity fluctuates randomly above and below the separatrix line;  in (B, C) the mean-squared activity remains at only one side of the separatrix; in (D, E) the mean-squared activity fluctuates about the separatrix in sync with $\hat C_s(\tau)$.  }
    \label{fig:sample-trajectories}
\end{figure*}

In particular, in any single network instance 
the {\it instantaneous} mean-squared activity  after a transitory time $t_0$
\be
\hat{q}_s  (t_0 + \tau) = \frac{1}{N}\sum_i x_{i,s}^2(t_0 + \tau)
\ee
fluctuates in time ($\tau$) about the stationary value $\hat C_s(0)$ (unless the system is at a fixed-point steady state). Therefore, in the case in which $\hat C_s(0)$ is sufficiently close to $C_0^*$, such fluctuations can effectively move the system above and below the separatrix as illustrated in Fig.\ref{fig:potential}. 
This leads to a motion that is highly irregular and sensitive to perturbations and small changes in initial conditions, as shown in Fig.\ref{fig:sample-trajectories}(A).  As a result, the motion turns out to be chaotic. More specifically: the time-lagged correlator, computed as a sliding time window average over a single dynamical trajectory \footnote{For an ergodic system, where time averages are equivalent to ensemble averages, such quantity   should equate the ensemble-averaged correlator $C(\tau)$, in the thermodynamic limit, if the system is at stationarity.},
\be
\hat C_s(\tau)=\frac{1}{t_m}\sum_{t=t_0}^{t_m}\frac{1}{N}\sum_i x_{i,s}(t)x_{i,s}(t+\tau)
\label{eq:Cs_tau}
\ee
does not exhibit stationary 
behaviour, but an irregular motion which is not time-translation invariant (TTI), i.e. it retains a dependence on $t_0$, see Fig.\ref{fig:sample-trajectories}(A), bottom.
 Our analysis in App. \ref{sec:stability-SS} shows that, in the thermodynamic limit, stationary solutions are unstable against perturbations that break TTI, suggesting that such non-TTI steady states are selected in {\it large} networks. 

On the other hand, when $C_s(0)$ is far from $C_0^*$, small fluctuations cannot shift the system across the separatrix and the resulting motion is periodic, as illustrated in Fig.\ref{fig:sample-trajectories}(B, C). 
As a consequence, in {\it finite} networks of size $N$, the system exhibits periodic dynamics in wide regions of the phase diagram, where $C_s(0)$ is 
far from $C_0^*$ (see Fig.\ref{fig:non_fixed_points}(E)). In addition, Fig.\ref{fig:non_fixed_points}(E) shows that, at finite $N$, there is a finite probability that the system exhibits fixed-point dynamics, at sufficiently high values of $1/gJ$. 
\begin{figure*}[tbh]
    \centering
    \includegraphics[width=1.0 \linewidth]{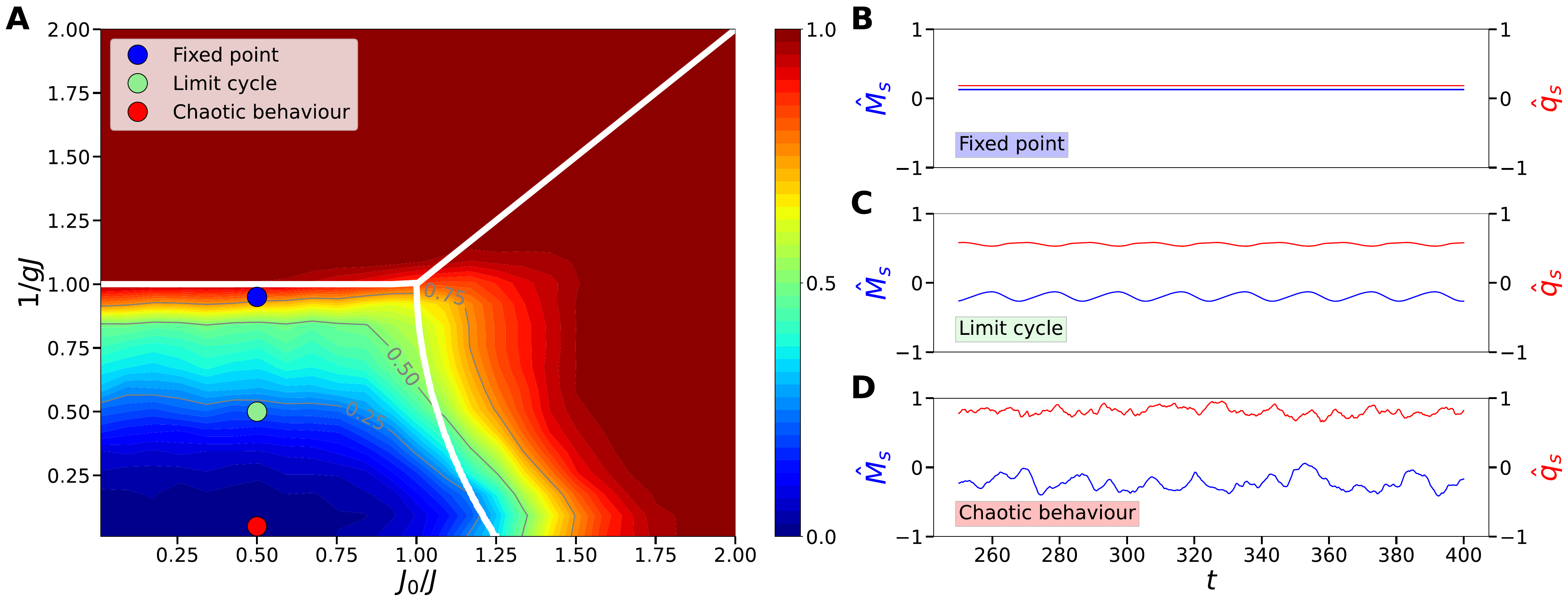}
    \caption{Probability of finding a fixed-point solution
    in the \textbf{uncorrelated} ($\gamma = 0$) and \textbf{noiseless} ($\sigma^2 = 0$) case in simulations with networks of size $N = 100$. {\bf Panel (A)}: Heat map of the PDF of finding a fixed-point solution for $S = 1000$ different realizations of the microscopic dynamics. Such a probability drops as $1/gJ$ is lowered, in agreement with Fig.\ref{fig:non_fixed_points}, showing that the probability that $C(0)=q$ vanishes at low values of $1/gJ$. In such a region
    a non-fixed-point solution characterized by chaotic behaviour emerges. {\bf Panels (B, C, D)}:  Typical trajectories observed in simulations at different values of the parameters $J_0/J, 1/gJ$ as indicated by circles on (A). Additional studies reveal 
    that, in the thermodynamic limit, chaotic solutions are the only
    stable solutions throughout the whole spin-glass phase. }  
    \label{fig:chaos_2}
\end{figure*}

\begin{figure*}[tbh]
    \centering
    \includegraphics[width=1.0 \linewidth]{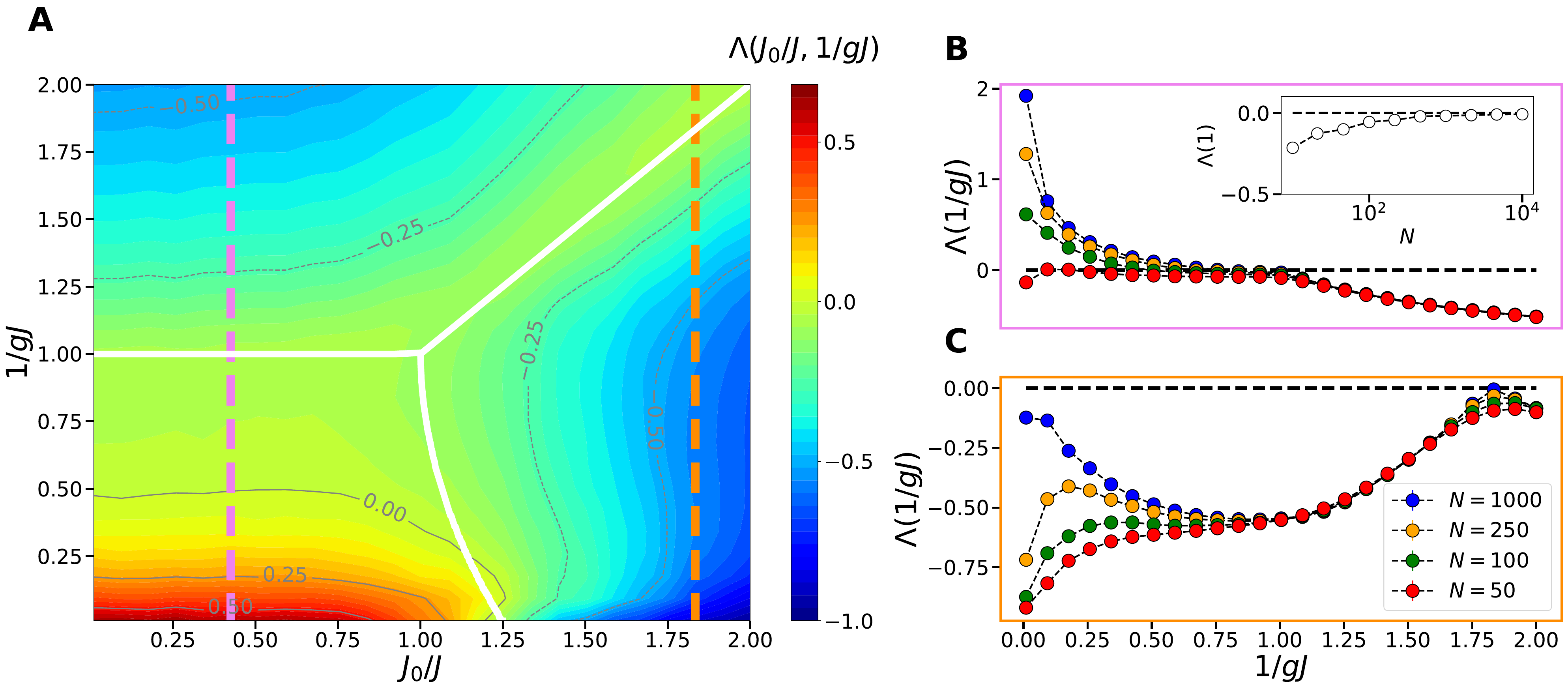}
    \caption{\textbf{Panel (A)}: Largest Lyapunov exponent (LLE) $\Lambda$ versus $J_0/J$, $1/gJ$
    as mesured in simulations for the \textbf{uncorrelated} ($\gamma = 0$) and \textbf{noiseless} ($\sigma^2 = 0$) case for networks of size $N = 100$ and $S = 1000$ realizations. Chaotic behavior emerges below the isocline $0.00$. \textbf{Panel (B, C)}: $\Lambda$ versus $1/gJ$ for different values of $N$, as shown in the legend, and $2$ different values of $J_0/J$, as indicated by the dashed lines on the left panel ((B) and (C) panels correspond to the left and right lines, respectively). At the smaller value of $J_0/J$ (panel (B)) there exists a critical value $1/g^\star J$ below which the LLE  becomes positive; such a value tends to $1$ as $N$ increases as show in the inset; in particular, $\Lambda(1/gJ = 1)$ as a function of $N$ converges asymptotically to 0. For the larger value of $J_0/J$ (panel (C)) $\Lambda$ is negative for all $1/gJ$ 
    and remains so by increasing network size; thus, there is no chaos in the ferromagnetic phase.} 
    \label{fig:chaos}
\end{figure*}

To illustrate all this, in Fig.\ref{fig:chaos_2} we show a heat map of the probability of finding fixed-point solutions, versus $J_0/J, 1/gJ$, in a system of (relatively small) size $N=100$, as well as typical trajectories exhibited by the system at representative points of the phase diagram. Chaotic trajectories are observed only at sufficiently low values of $1/gJ$, where the probability distribution function of $C(0)$ becomes peaked around $C_0^\star$, as shown in 
Fig.\ref{fig:non_fixed_points}. 

Furthermore, we note that, while chaotic trajectories seem to arise, in finite networks, from fluctuations around the separatrix, this does not seem a sufficient condition to have chaotic motion. For example, when fluctuations of the mean-squared activity around the separatrix value are periodic and in sync with the natural oscillations of the two-time correlator, as shown in Fig.\ref{fig:sample-trajectories}(D, E), the system is observed to settle on a periodic trajectory.

In this regard, the dynamics of the two-time correlator resembles that of a forced oscillator, where the external ``force" is given by the fluctuations of the initial condition. In particular, if the driving force is periodic and it has the same frequency as the natural frequency of the oscillator, the system
oscillates at the frequency of the applied force as shown in Fig.\ref{fig:sample-trajectories}(B, C). If the frequency of the applied force is different from the natural frequency of the oscillator, the system may exhibit sub-harmonic oscillations i.e. a periodic motion with a frequency that is a fractional multiple of the input frequency, as shown in Fig.\ref{fig:sample-trajectories}(C, E).

As discussed above, for large $N$, $C(0)$ is peaked around $C_0^\star$, hence the scenario discussed in Fig. \ref{fig:sample-trajectories}(B, C) cannot arise, as dynamical fluctuations of the mean-squared activity about its time-averaged value move the system above and below the separatrix curve. However, periodic trajectories as described in Fig.\ref{fig:sample-trajectories} (D, E), can in principle still arise. In order to determine whether periodic motion survives for large $N$, it is therefore important to establish whether such trajectories remain stable in the thermodynamic limit. Our analysis in App. \ref{sec:stability-SS} suggests that they lose stability for large $N$, as stationary trajectories in the spin-glass region are unstable, in the thermodynamic limit, against perturbations that break TTI. 

Numerical evaluation of the largest Lyapunov exponent (LLE) $\Lambda$, in simulated trajectories of the system (see App.\ref{app:LLE}) shows that, upon increasing $N$, only chaotic orbits prevail inside the spin-glass region. Fig.\ref{fig:chaos} (A) shows that at small system sizes, $\Lambda$ is positive for small values of $1/gJ$ within the spin-glass region, and it decreases to zero as $1/gJ$ increases.  However, as the system size increases, the value of $\Lambda$ at $1/gJ=1$ increases and eventually vanishes (see inset of Fig.\ref{fig:chaos}, panel B), suggesting that, in this limit, $\Lambda$ remains positive for any $1/gJ<1$. On the other hand, in the ferromagnetic region 
$\Lambda$ remains negative for any value of 
$1/gJ$, 
see Fig.\ref{fig:chaos} (C). The inset in panel (B) indicates that the approach to zero of $\Lambda$ at $1/gJ=1$ is slow, suggesting that in large but finite systems, where $C(0)$ has already converged to the separatrix value $C_0^\star$, periodic motion as described in Fig.\ref{fig:sample-trajectories} (D, E) can still be observed. 
However, as explained earlier, the stability analysis of stationary trajectories carried out in App.\ref{sec:stability-SS} suggests that such a periodic motion eventually loses its stability.

\begin{figure*}[tbh]
    \centering
    \includegraphics[width=1 \linewidth] {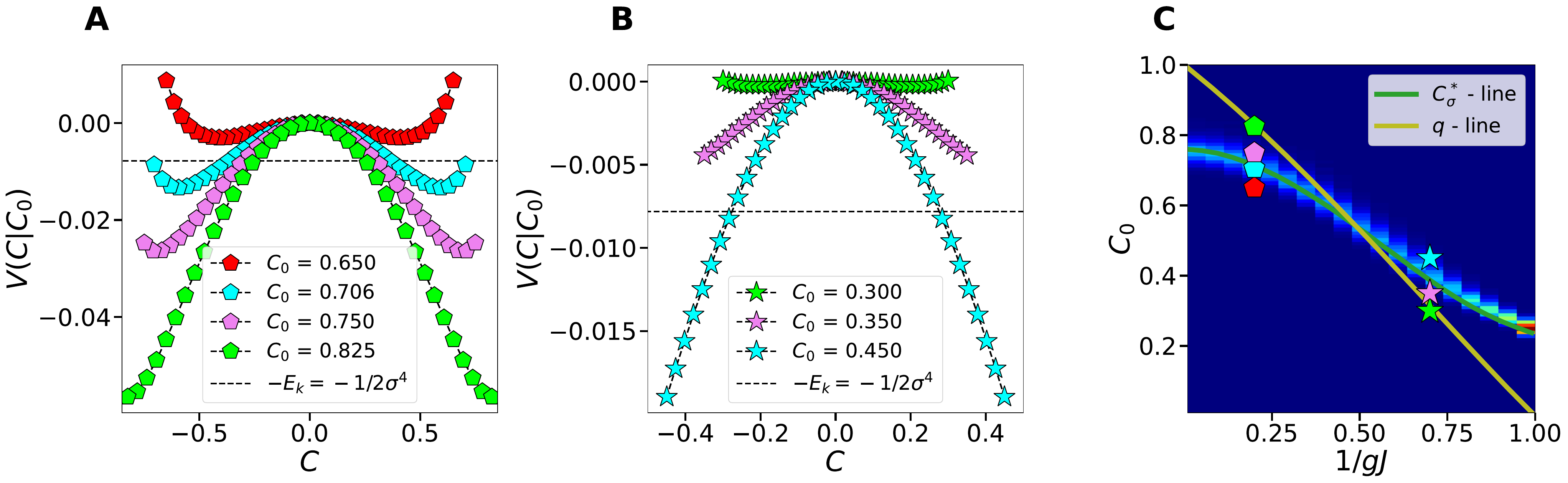}
    \caption{Analysis of the motion in the effective potential $V(C|C(0),M)$ 
    for the \textbf{uncorrelated} ($\gamma = 0$) and \textbf{noisy} ($\sigma^2 \neq 0$) case, for $M\!=\!0$ (spin-glass phase, with $J_0/J = 0.5$). 
    \textbf{Panels (A, B)}: Potential $V(C | C_0)$ as a function of $C$ for fixed $1/gJ = 0.2$ and $1/gJ = 0.7$, respectively; plotted in the range $C\in[-C(0),C(0)]$ for different values of $C_0$ (as shown in the legend, with a color code). The dashed black line shows the negative of the initial kinetic energy, $-1/2\, \sigma^4$, arising from the noise. When this value is equal to the initial potential energy, the total energy is zero, corresponding to motion on the separatrix (blue curve in A).
    \textbf{Panel (C)}: Heat map of the probability distribution function (PDF)  $P(C_0)$ of finding a given value of $C(0)$ in the SG phase, measured from $S = 1000$ simulations of the microscopic dynamics averaged over a time window $t_m = 2000$ for system size $N = 100$. Stars and pentagons show the values of $1/gJ$ and $C_0$ used in panel (B) and (C), respectively. Solid lines show the values of $C(0)$ corresponding to $C_\sigma^*$ (green) as defined in Eq.(\ref{eq:C_sigma_star}) and $q$ (yellow). The intersection between both lines defines the critical value $1/g^* J$ at which chaos emerges in the infinite-size limit (see Sec. \ref{sec:stability-SS}). } 
    \label{fig:noise_C0}
\end{figure*}

In summary, we have found two types of criticality, one SG like (type-II), and the other ferromagnetic like (type-I). All across the SG phase the system exhibits chaotic behavior in the thermodynamic limit, while in finite networks fixed points and periodic oscillations may emerge, especially close to the transition line.  

\subsection{Switching on fluctuations:\\
Noisy dynamics ($\sigma\neq 0$) with uncorrelated interactions ($\gamma=0$)}
\label{sec:sn0-g0}

\begin{figure*}[tbh]
    \centering
    \includegraphics[width=0.65 \linewidth] {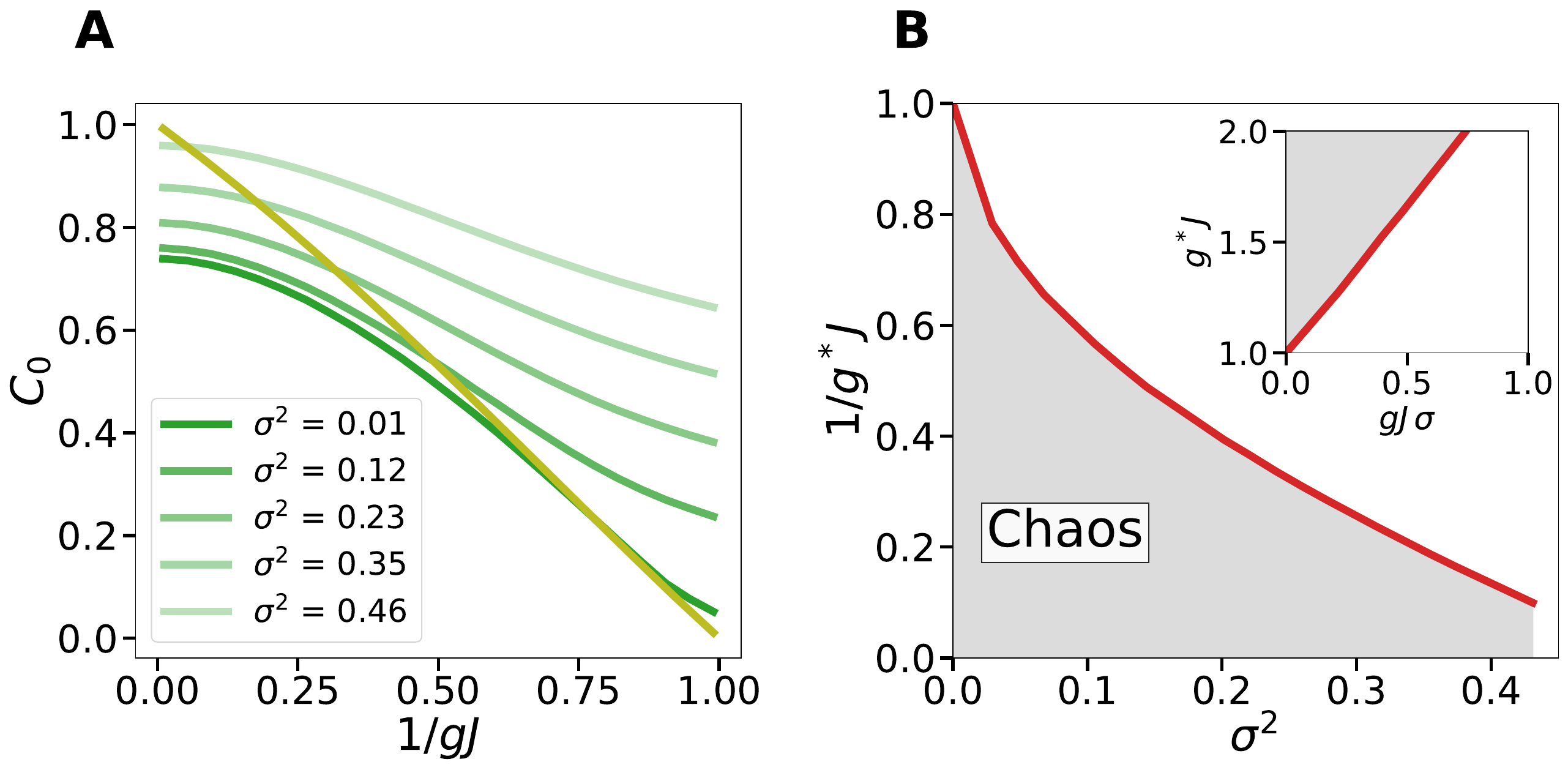}
    \caption{ Theoretical prediction of the critical line for the \textbf{uncorrelated} ($\gamma = 0$) and \textbf{noisy} system ($\sigma^2 \neq 0$). \textbf{Panel (A)}: $C_\sigma^\star$, as defined in Eq.(\ref{eq:Cstar-noise}), as a function of $1/gJ$ for different values of $\sigma^2$ (green lines) plotted together with  the solution of Eq.(\ref{eq:C_sigma_star}) as a function of $1/gJ$ (yellow line). The intersection between each green line and the yellow one define the critical value $1/g^* J$ at which chaos emerges in the infinite-size limit for each value of $\sigma$. \textbf{Panel (B)}: Critical value $1/g^\star J$ as a function of $\sigma^2$. As the noise strength $\sigma^2$ increases, the chaotic region delimited by $1/g^* J$ (grey region) shrinks, eventually disappearing when the noise amplitude is sufficiently large. In the inset, we plot the critical value $g^* J$ as a function of the "scaled" $\sigma' = \sigma \, g J$, where our model becomes equivalent to the SCS (see App.  \ref{app:equivalence}). Results are consistent with the critical line derived in \cite{Helias2018}.  }  
    \label{fig:noise_critical_line}
\end{figure*}


In this Section we analyze the effect of additional sources of external variability ---in particular the presence of an external source of noise, i.e.  $\sigma\neq 0$--- on the firing-rate model for uncorrelated networks ($\gamma=0$). 
In this case, there cannot be any fixed-point solution, as the rates $x_i(t)$ fluctuate stochastically. Non-fixed point steady states are characterised by the same magnetization $M$ as in the noiseless dynamics, given by Eq.\eqref{eq:M-self-asym-SS}, while the equation for the correlator needs to be modified to (see SM, Sec. \ref{app:C-dynamics-SS})
\bea
\partial_\tau^2 C=-\frac{\partial V(C|C(0),M)}{\partial C}-2\sigma^2 \delta(\tau).
\label{eq:motion_in_Vt}
\eea
Integrating over $\tau \in [0,\epsilon]$, with $\epsilon\ll 1$ leads to
\be
\dot C(\epsilon)=\dot C(0)-\sigma^2 +{\mathcal O}(\epsilon)
\ee
and recalling that $C(\tau)=C(-\tau)$ implies $\dot C(0)=0$, one gets for $\epsilon \to 0^+$ that the initial velocity is $\dot C(0^+)=-\sigma^2$, as previously found in \cite{Helias2018}. From now on, for the sake of simplicity, we focus on the case $M=0$. 

\begin{figure*}[tbh]
    \centering
    \includegraphics[width=0.7 \linewidth]{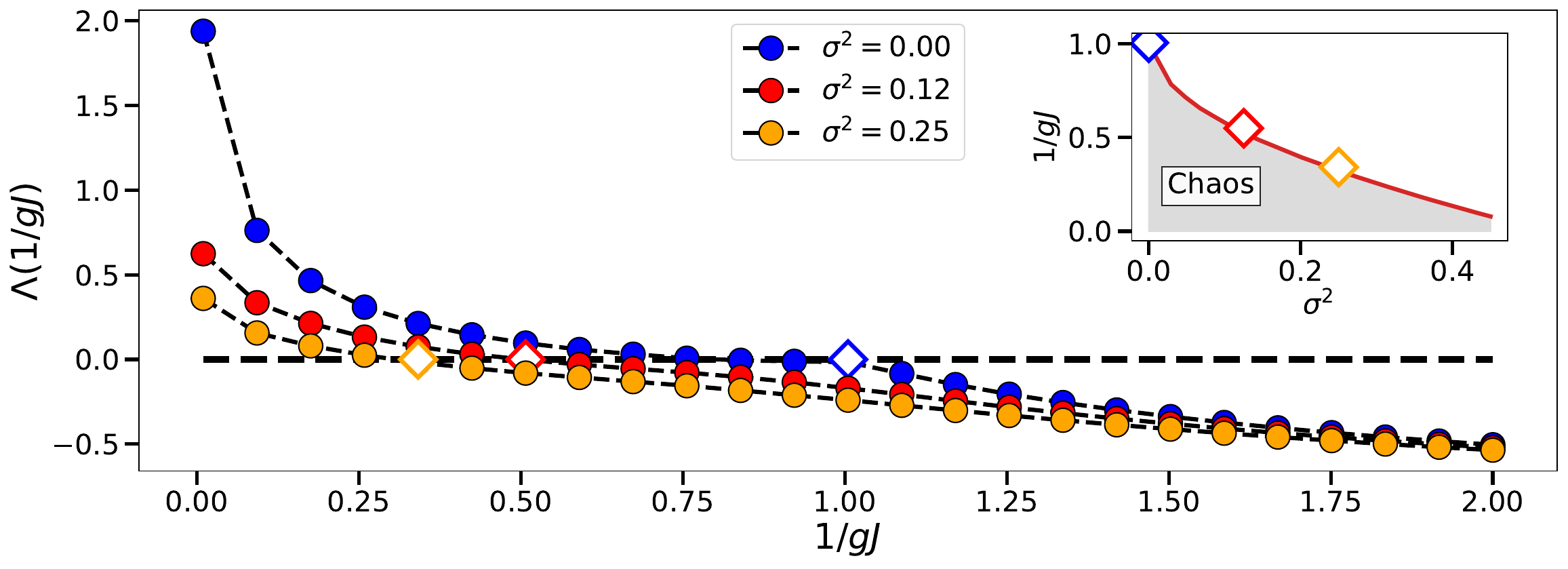}
    \caption{For the \textbf{uncorrelated} ($\gamma = 0$) system, the LLE $\Lambda$ obtained from numerical simulations is shown as a function of $1/gJ$, for $J_0/J = 0.5$ and different noise levels $\sigma^2 = 0, 0.12$ and $0.25$, as shown in the legend. Results are averaged over $S=100$ realizations 
    and system size $N=1000$. In the inset, we reproduce Fig.\ref{fig:noise_critical_line}(B), adding the value $1/g^* J$ at which $\Lambda$ becomes positive (white diamonds).}  
    \label{fig:noise_lyapunov}
\end{figure*}

Let us remark that, as noticed earlier,  $V'(C(0)|C(0),0)<0$ for $C(0)>q$, so that this led to un-physical solutions in the {\it noiseless} case, where $\dot C(0)=0$. However, the situation is different in the presence of noise; given that the initial velocity is {\it negative}, a physical solution can {\it still} be obtained if the total energy is equal to zero, which corresponds to the separatrix curve, i.e. asymptotic motion towards $C=0$ (see Fig.\ref{fig:noise_C0}(B)). This occurs for the initial condition $C(0)=C_\sigma^*$ such that
\begin{equation}
V(C_\sigma^*|C_\sigma^*,0)=-\frac{1}{2}\sigma^4.
\label{eq:Cstar-noise}
\end{equation}
On the other hand, {\it any} $C(0)\leq q$ is physically plausible as it leads to bounded motion. To ascertain the value of $C(0)$ that is dynamically selected by the system, we plot 
in Fig.\ref{fig:noise_C0}(C) the distribution $P(C_0)$, defined in Eq.\eqref{eq:PDF_Cs}, computed from $S=1000$ simulations of the microscopic dynamics, 
versus $1/gJ$, as a heat map, as well as the value of $C_\sigma^*$ and $q$, as defined in Eq.\eqref{eq:Cstar-noise} and Eq.\eqref{eq:q-self-asym}, respectively.
Results show that 
$P(C_0)$ concentrates on the curve $C(0)=C_\sigma^*$. 
As we explained above, while for $C(0)>q$ this is the {\it only physical} solution, for $C(0)\leq q$, there are in principle many possible solutions. A stability analysis of steady-state solutions (see App. \ref{sec:stability-SS}) shows however that such solutions are all unstable, except for $C(0) = C_\sigma^*$, which is \textit{marginally} stable. Then, the system selects the one laying on the separatrix curve, which delimits different possible (unstable) motions.
As any fluctuation leads to an instability, chaotic behaviour is expected to emerge in single instances for $C_\sigma^*<q$. The critical line $C_\sigma^*=q$ is thus predicted to mark the transition to chaos. In Fig.\ref{fig:noise_critical_line}(A) we plot $q$ and $C_\sigma^*$ versus $1/gJ$, for different values of $\sigma$. The intersection between $C_\sigma^*$ and $q$, obtained from
\be
C_\sigma^\star=\int \D\psi \, \tanh^2(gJ\sqrt{C_\sigma^\star}\psi)
\label{eq:C_sigma_star}
\ee
gives a curve in the $(\sigma^2,1/gJ)$-plane plotted in Fig.\ref{fig:noise_critical_line}(B). 
A similar condition to Eq.\eqref{eq:C_sigma_star} has been derived in \cite{Helias2018} for the model originally defined by SCS in \cite{Crisanti88}.
Let us remark that the variant of the model we consider here has a narrower chaotic region than the original model, in spite of the fact that their behaviour is identical in the absence of noise. This observation can be rationalised by a simple argument (see App. \ref{app:equivalence}) which shows that in our variant of the model the noise strength $\sigma^2$ is effectively amplified by a factor $(gJ)^2$, which is larger than $1$, in the spin glass region. In the inset of  Fig.\ref{fig:noise_critical_line}(B) we show the equivalent curve $g^* J$ as a function of $\sigma \, g^* J$, for fixed $J$, recovering precisely the result obtained in \cite{Helias2018}. 

The emergence of chaos in the noisy dynamics is confirmed by numerical analysis of the LLE, shown in Fig.\ref{fig:noise_lyapunov}. For $J_0/J = 0.5$, the LLE $\Lambda$ is plotted as a function of $1/gJ$ for different values of $\sigma^2$ and fixed $N = 1000$. The critical values $1/g^* J$ at which the $\Lambda$ becomes positive, i.e. $\Lambda(1/g^* J) = 0$, are plotted in the inset, which are in agreement with the critical line defined in Eq.(\ref{eq:C_sigma_star}) (solid, red line).

Thus, the size of the chaotic phase is reduced by the presence of external noise, so that the boundary of the chaotic phase is shifted towards larger coupling values, in agreement with previously reported results.

\subsection{Switching on correlations
($\gamma \neq 0$)} 
\label{sec:s0-gn0} 

In this section, we analyse the dynamics
when interactions are correlated ($\gamma\neq 0$).
Any degree of correlations introduces a so-called ``retarded'' self-interaction in the equation for the effective single-neuron Eq.\eqref{eq:single-neuron} i.e. a term non-local in time, which makes the dynamics non-Markovian and the analysis considerably harder. For this reason, here, we restrict our analysis to fixed-point solutions of the noiseless dynamics ($\sigma=0$). 

\begin{figure*}[t]
    \centering
    \includegraphics[width=1.0  \linewidth]{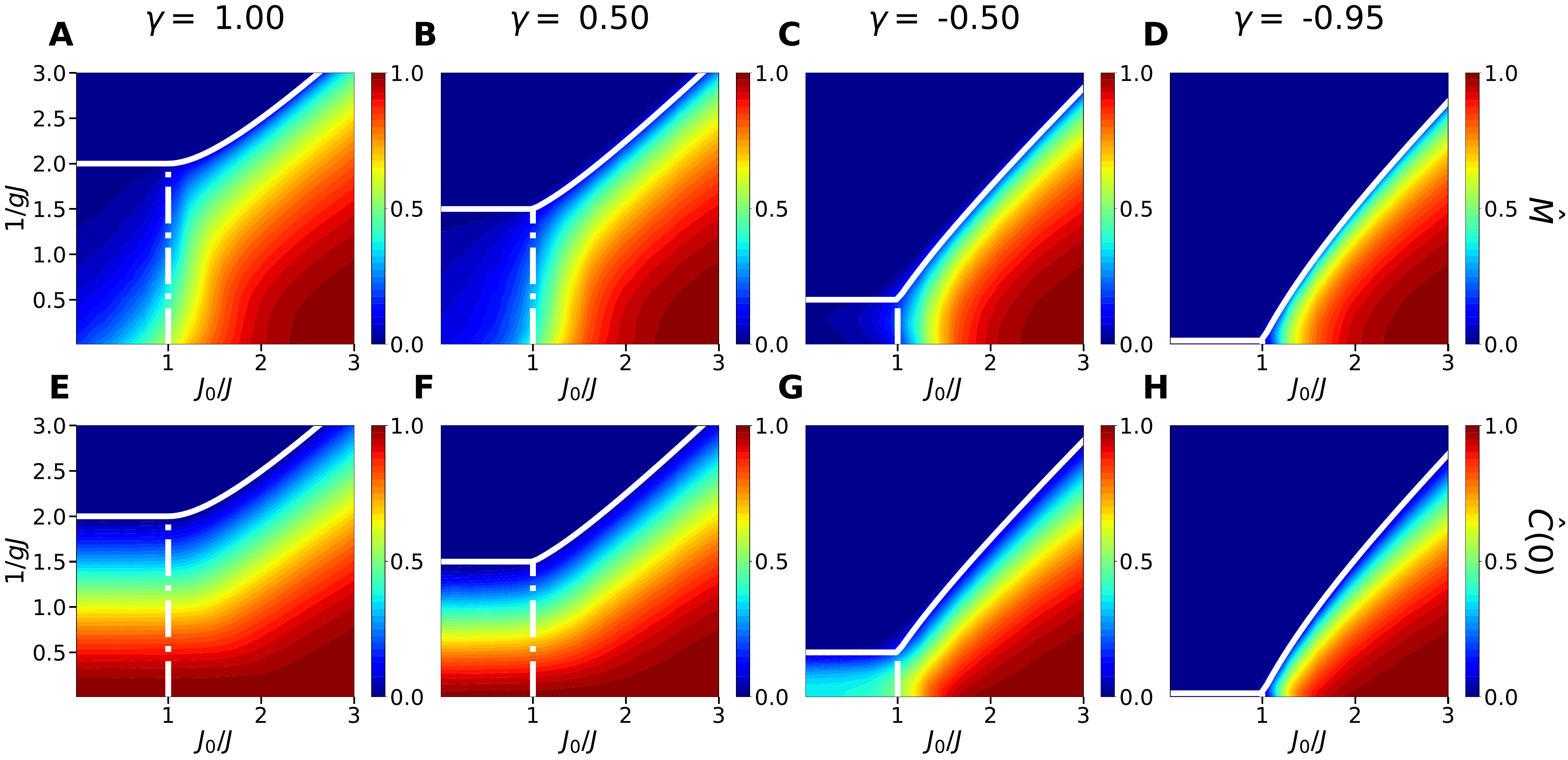}
\caption{Phase diagram emerging for simulations for networks with \textbf{correlated} couplings ($\gamma \neq 0$) and \textbf{noiseless} dynamics ($\sigma^2 = 0$). \textbf{Panel (A, B, C, D)}: Heat map of the time-averaged mean activity $\hat{M}$ for $\gamma = 1, 0.5, -0.5$ and $-0.95$, respectively. \textbf{Panels (E, F, G, H)}: Heat map of the time-averaged mean-squared activity $\hat{C}(0)$ for $\gamma = 1, 0.5, -0.5$ and $-0.95$, respectively. Solid, white lines show the theoretical prediction Eq.(\ref{eq:critical-line-gamma}), while dotted, white lines show the curve $J_0/J=1$, for visual aid. Each panel has been obtained integrating the microscopic dynamics for a system size $N = 100$ and averaging over $S = 1000$ different realizations. }  
    \label{fig. parameter_space symmetric}
\end{figure*}

In the fixed-point regime, the response function is time-translation invariant $R(t,t')=R(t-t')$ and $C(t,t')$ becomes independent of $t,t'$, so that it can be written as $q=C(t,t')$. As before, each realization of $\phi(t)$ becomes a static Gaussian random variable with zero mean and variance $g^2J^2 q$.  Setting $\phi=gJ\sqrt{q}\psi$, where $\psi$ is a zero-average random variable with unit variance, the fixed point of Eq.\eqref{eq:single-neuron} satisfies, for each realization of $\psi$, 
\be
x(\psi)=\tanh [J_0g M+\gamma J^2 g^2 \chi x(\psi)+gJ\sqrt{q}\psi]
\label{eq:x-self-psi}
\ee
where $\chi=\int_0^\infty\,d\tau R(\tau)$ is the integrated response, which is found from Eq.\eqref{eq:Delta_C} as
\bea
\chi&=& \left\langle \frac{\partial x(\phi)}{\partial \phi} \right\rangle\equiv
\frac{1}{gJ\sqrt{q}}
\left\langle \frac{\partial x(\psi)}{\partial \psi}
\right \rangle.
\label{eq:chi-phi-psi}
\eea

Averaging Eq.\eqref{eq:x-self-psi} over $\psi$, we obtain
\bea
M&=&\bra x(\psi)\ket 
\\
&=&\int D\psi\tanh [J_0g M+\gamma J^2 g^2 \chi x(\psi)
+gJ\sqrt{q}\psi],
\nonumber
\label{eq:M-self-psi}
\eea
and similarly 
\bea
q&=&\bra x^2(\psi)\ket \\
&=&\int D\psi\tanh^2 [J_0g M+\gamma J^2 g^2 \chi x(\psi)
+gJ\sqrt{q}\psi].
\nonumber
\label{eq:q-self-psi}
\eea
As before, for $g=0$ the system is in the paramagnetic phase $M=0$, $q=0$ and we expect this solution to remain stable below a critical value of $g$.
However, in contrast with the case $\gamma=0$, the equations for the moments $M$ and $q$ do not close, as they retain a dependence on the realizations $x$ of the stochastic process. Nevertheless, progress can be made close to the paramagnetic instability, where $\tanh(x)$ can be linearised and equation closure can be achieved.  In App. \ref{app:FP_diagram_gamma} we show that the critical line where the paramagnetic phase becomes unstable can be calculated explicitly as
\be
\frac{1}{gJ}={\rm max}\Big(\gamma+1, \frac{J_0}{J}+\gamma\frac{J}{J_0}\Big)
\label{eq:critical-line-gamma}
\ee
Eq.\eqref{eq:critical-line-gamma} retrieves the line Eq.\eqref{eq:bifurcation_Linear_0} for uncorrelated coupling ($\gamma=0$) and  
Eq.\eqref{eq:bifurcation_Linear_1} for reciprocal interactions ($\gamma=1$), and it is consistent with results from random matrix theory.  
In particular, the first term on the RHS of Eq.\eqref{eq:critical-line-gamma} recovers the expression for the largest eigenvalue in the spectral distribution of partially symmetric Gaussian matrices with zero mean derived in \cite{sommers_spectrum_1988}, and the second term recovers the expression for the outlier eigenvalue of partially symmetric Gaussian matrices with non-vanishing mean derived in \cite{orourke_low_2014}.
As noted in \cite{CureNeri23}, the largest eigenvalue of the bulk and the outlier are each given by the sum of their values at $\gamma=0$ and a term which is proportional to $\gamma$. In the context of our analysis, this term can be rationalised as the additional contribution that $M$ and $q$ receive from the susceptibility (i.e. the integrated response), around the instability line. This relation shows that positively correlated interactions ($\gamma>0$)  enlarge the regions of ferromagnetic and spin-glass order, with respect to the case of uncorrelated interactions ($\gamma=0$), while negatively correlated interactions ($\gamma<0$) shrink the extension of the ordered phases, with the spin-glass region disappearing at $\gamma=-1$.

Finally, note that, owing to the non-linear and non-closed nature of equations, analyses away from criticality cannot be carried out exactly. Thus, we resort to numerical simulations.

In Fig.\ref{fig. parameter_space symmetric} we show 
a heat map of the time averaged $\hat M$ and $\hat C(0)$ (computed from numerical simulations) in the parameter space $J_0/J,1/gJ$,
together with the theoretically determined critical line Eq.(\ref{eq:critical-line-gamma}), for different values of $\gamma$ (computed from the stability condition of fixed-point solutions). Observe that, even if non-fixed-point solutions are expected to appear in the ordered phase when interactions are non-symmetric (i.e. $\gamma<1$), the theoretical lines obtained for fixed points are seen to capture well the actual phase-transitions for all values of  $\gamma$. 
The figure clearly confirms the theoretical analyses: the spin-glass phase is enlarged for positively correlated couplings and progressively shrinks for anticorrelated ones (and disappears in the limit case $\gamma=-1$) and the theoretical lines  correctly reproduce the actual phase transition in all cases.

\begin{figure*}[tbh]
    \centering
\includegraphics[width=0.7  \linewidth]{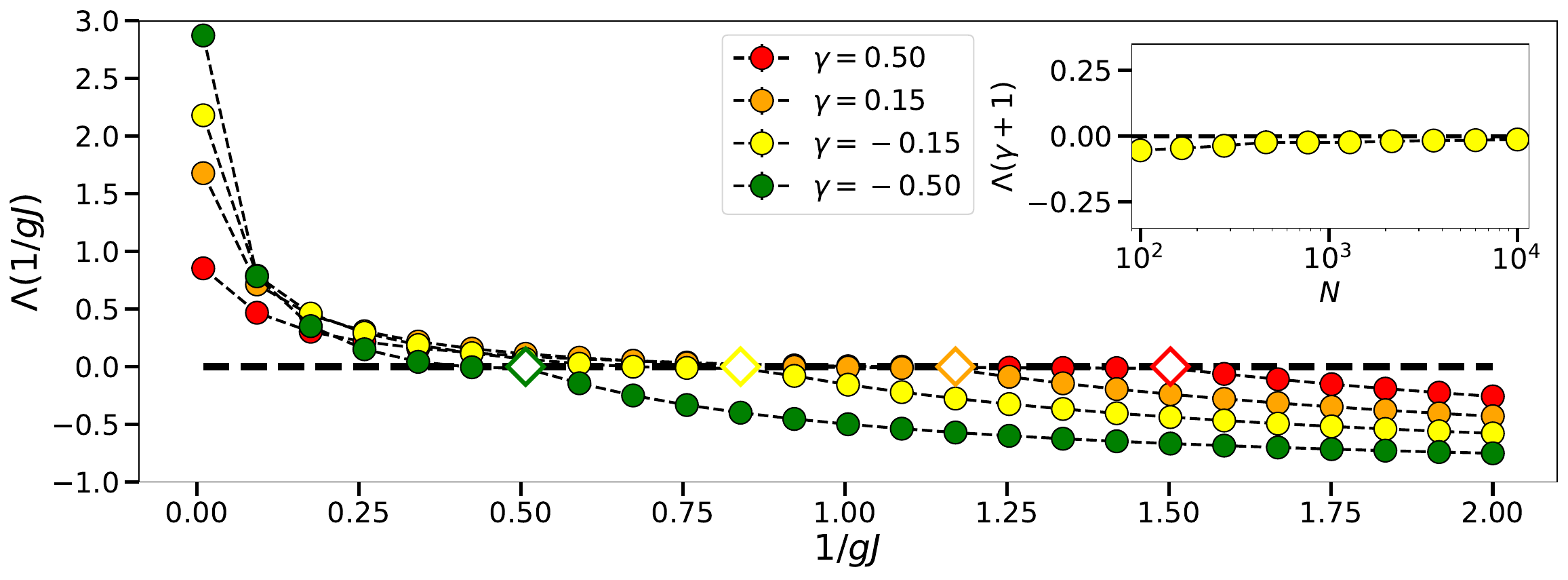}
    \caption{For \textbf{correlated} ($\gamma \neq 0$) and \textbf{noiseless} dynamics ($\sigma^2 = 0$), Largest Lyapunov exponents obtained from simulations of the system, with size $N = 1000$, averaged over $S = 100$ realizations, for $J_0/J = 0.5$ and different values of $\gamma$, as shown in the legend. The point at which the LLE becomes positive is depicted as a white diamond. The inset shows the value of $\Lambda$ at $1/g J = \gamma + 1$, for $\gamma = 0.15$, as a function of $N$. }  
    \label{fig:lyapunov-asym-N}
\end{figure*}

Also, in Fig.\ref{fig:lyapunov-asym-N} we show results for the LLE as a function of the coupling strength ($1/gJ$), computed for different values of $\gamma$. The curves show that, for all values of $\gamma$, the exponents become positive below some critical value (marked with diamond symbol) that is consistent with the theoretical prediction $1/g_cJ=1+\gamma$. Finite-size effects are quantified in the inset (for $\gamma=-0.15$).

Some remarks are in order:
\begin{itemize}
\item The critical value (in terms of $1/gJ$) decreases as $\gamma$ is reduced, so that it is maximal for $\gamma=1$ (in agreement with the theoretical prediction).

\item Smaller values of $\gamma$ have larger LLEs in the limit of large coupling values (small $1/gJ$'s), so that, in this regime, the dynamics becomes more chaotic for anti-correlated couplings than for positively correlated ones.

\item Indeed, as the value of $\gamma$ approaches $1$, the LLE becomes very close to $0$ all across the spin-glass phase, so that, actually, the dynamics becomes marginally stable in such a limit.  This last result resembles the
recent observation  of a phase of marginal stability in fully-correlated (i.e. symmetric) random networks in models for theoretical ecology \cite{BiroliCammarota18}. Dynamics in such region can be regarded as ``at the edge of chaos", as defined in other contexts see e.g. \cite{gene-Thurner08, Thurner2010,Morales1,Morales2}.
\end{itemize}

In summary, the introduction of correlations in the couplings quantitatively modifies the shape of the phase diagram with respect to the uncorrelated case: the spin-glass region is enlarged for positive correlations and shrinks for anticorrelated ones. On the other hand, the ``strength" of chaos is larger for anticorrelations and tends to be reduced by positive correlations, leading to marginal stability in the limit of perfectly correlated couplings.

\section{Discussion and Conclusions}
\label{sec:conclusions}

In this study we have considered a simple neural network rate model with random Gaussian interactions having non-vanishing mean and possible correlations, i.e. non-reciprocal interactions. We have analyzed the dynamics, averaged over the network ensemble in the thermodynamic limit, using  a path integral formalism. These analytical studies have been complemented with extensive computational simulations for finite networks.

In general, much as in the physics of standard (reciprocal) disordered systems, fixed-point solutions are characterised by two order parameters, namely the mean ($M$) and the variance ($q$) of the neural activity. Thus,  two different types of criticality emerge depending on whether the mean becomes non-zero (type-I) or not (type-II) at the transition, as the strength of the coupling is increased.  Such criticalities correspond to the transition of the system from the disordered phase to either a ferromagnetic (type-I) or spin-glass (type-II) phase, respectively.

Our analyses show that as soon as some degree of non-reciprocity is switched on ---i.e., for any value of the coupling correlations, from nearly symmetric to anti-symmetric, $ -1 < \gamma < 1$--- chaotic behavior emerges within the spin-glass phase ($M=0$, $q\neq0$).
 We have also shown that
 the size of the chaotic phase is reduced when external noise is added, so that ``noise suppresses chaos", as it had already been found in previous works \cite{Helias2018}. Furthermore, we have shown how the presence of correlations (i.e. some degree of reciprocity) in the couplings changes the phase diagram: negative correlations shrink the spin-glass phase (where chaotic dynamics emerges) while positive correlations enlarge it. On the other hand, negative correlations increase the strength of chaos while positive correlations reduce it, with chaos disappearing for symmetric couplings, where equilibrium is attained and the whole spin-glass phase exhibits marginal stability (much as in \cite{BiroliCammarota18}). 

Chaotic phases for networks of non-reciprocal couplings ---which define an inherent non-equilibrium problem--- exhibit intriguing analogies with equilibrium spin-glass phases. For example, we have shown that in the chaotic phase, the dynamical equation of the global two-time correlation function allows for {\it multiple steady states}, including fixed points and periodic orbits. Our careful analyses of the dynamics reveal that the system {\it dynamically} selects the steady state that corresponds to motion along the {\it separatrix} curve, which delimits the basins of attractions of different types of solutions. This is reminiscent of the behaviour of equilibrium systems in spin-glass (fRSB) phases, where a multitude of steady states emerges and the system is observed to select {\it marginally stable} states, i.e. saddles in the free-energy landscape, that are linked together by flat directions. Our analyses also suggest that chaotic motion is the manifestation, at the level of {\it single network instances}, of an ensemble-averaged dynamics that lies on the separatrix curve delimiting different possible steady states. Such a motion is not time-translation invariant (TTI), similarly to ageing dynamics in spin-glass phases.

This last consideration prompted us to introduce a method to detect the onset of chaotic behaviour, which is based on a linear stability analysis of the steady-state solutions against perturbations that break TTI. Our method is 
consistent with the two-replica approach described in  
\cite{Crisanti18, Helias2018}, but it is considerably simpler, and it shows that chaotic behaviour emerges whenever the equation for the global correlation function allows for multiple steady-states.  
The sensitive dependence on initial conditions in the chaotic phase means that two "replicas", i.e. two copies of the system with the same disorder and initialised in different configurations, display different dynamical behaviour, akin to replicas of an equilibrium system ending up in different states.  Furthermore, we find that ---while the global correlator has multiple steady states in the chaotic phase--- both a signal (i.e. a non-vanishing mean connectivity) and noise bring the system to a phase where there is only one steady state. This, again, mirrors what happens in equilibrium settings, where signal and thermal noise are known to induce transitions from the spin-glass phase to the ferromagnetic and paramagnetic phase respectively.

A theoretical prediction for the transition line between the spin-glass and the ferromagnetic region, for general values of the coupling asymmetry as well as a derivation of the asymptotic behaviour of the order parameters 
near criticality are  currently missing.  In particular, as we have seen, when interactions are not uncorrelated (or fully asymmetric), the equations for the time-dependent moments of the stochastic process do not close. This prevents an exact analysis away from the paramagnetic instability, to address the previous questions. 
However, one may be able to use closure schemes (e.g. Gaussian closure) as an approximation, to further characterize the phase diagram of fixed-point solutions. In addition, one may be able to get, within such approximation schemes, a closed equation for the two-time correlation at stationarity, that can be cast as a gradient-descent on a potential, and extend the analysis carried out for fully asymmetric couplings to interactions with arbitrary asymmetry. Another outlook could be adapting the approaches developed in \cite{Laffargue_2013, Laffargue_2015} to either numerically sample or calculate analytically fluctuations in the LLE and determine the model parameters where such fluctuations get atypically large, signaling the onset of chaos. 

Finally, it would be important to further investigate  whether chaotic dynamics and its links with spin-glass phases survive when different couplings, other than Gaussian, are chosen. For example, for applications in neuroscience, it would be interesting to carry out a similar analysis for couplings which satisfy Dale's rule \cite{Rajan} or are non-Gaussian 
\cite{Wardak22}, whereas the choice of {\it sparse} couplings may be relevant to applications in biology (e.g. gene regulatory networks \cite{gene-Thurner08, Torrisi_2020}) and theoretical ecology (e.g. Lotka-Volterra models \cite{BiroliCammarota18,Altieri-2021,Galla-2023}). 

\vspace{0.25cm}
{\bf{Acknowledgments:}} We acknowledge the Spanish Ministry and Agencia Estatal de investigaci{\'o}n (AEI) through Project of I+D+i
Ref. PID2020-113681GB-I00, financed by MICIN/AEI/10.13039/501100011033
and FEDER ``A way to make Europe”, as well as the Consejer{\'\i}a de
Conocimiento, Investigaci{\'o}n Universidad, Junta de Andaluc{\'\i}a
and European Regional Development Fund, Project reference P20-00173 for financial
support. AA thanks Carlos-I Institute  of Theoretical and Computational Physics, for the kind hospitality.

\onecolumngrid

\newpage

   \begin{center}
   {\bf{APPENDICES}} 
   \end{center}

\begin{appendix}

\section{On the equivalence between our model and the classic one (SCS)}
\label{app:equivalence}

In this section we show that the model originally introduced in \cite{Crisanti88} can be mapped to 
the variant we introduce here, when interactions are uncorrelated and zero-averaged, upon suitably scaling 
the noise strength. Multiplying Eq.\eqref{eq:dynamics-def},
\be
\dot x_i=-x_i+\tanh(g\sum_j J_{ij}x_j)+\xi_i,
\ee
times $gJ_{\ell i}$ and summing over $i$, we have, 
upon defining $y_\ell=g\sum_i J_{\ell i} x_i$ and 
$\phi_\ell=g\sum_i J_{\ell i}\xi_i$
\be
\dot y_\ell=-y_\ell+g\sum_i J_{\ell i}\tanh(y_i)+\phi_\ell
\ee
where
\begin{eqnarray}
\bra \phi_\ell(t) \phi_k(t')\ket &=&g^2\sum_{ij}J_{\ell i}J_{kj}\bra \xi_i(t)\xi_j(t')\ket=2g^2\sigma^2\sum_{ij}J_{\ell i}J_{kj}\delta_{ij}\delta(t-t')
\nonumber\\
&=&2g^2\sigma^2\sum_{i}J_{\ell i}J_{ki}\delta(t-t')
\end{eqnarray}
For uncorrelated, zero-mean Gaussian interactions $J_{ij}\sim {\mathcal N}(0,J^2/N)$, we have 
\be
\bra \phi_\ell(t) \phi_k(t')\ket=2g^2\sigma^2 J^2 \delta_{k\ell}\delta(t-t'),
\ee
hence the two models are fully equivalent when rescaling $\sigma$ with $gJ$. If, however, the $J_{ij}$'s are correlated or non-zero averaged, then the two models are no longer equivalent.

\section{Stability of steady-state solutions with uncorrelated interactions ($\gamma=0$)}
\label{sec:stability-SS}

In this section we study the linear stability of non-fixed points steady-states solutions.  
Away from stationarity, we can write the equation of motion for the two-time correlator $C(t,s)$ as 
\begin{equation}
(1+\partial_t)(1+\partial_s)C(t,s)  = F(C(t,s),C(t,t),C(s,s),C(s,t),M(t),M(s))
\label{eq:dot_Cts}
\end{equation}
where 
\be
F(C(t,s),C(t,t),C(s,s),M(t),M(s)) = 
\int
\mathcal{D}_{\phi, \phi'}(t,s) \; \tanh[J_0 g M(t)+\phi]\tanh[J_0 g M(s)+\phi']
\label{eq:F}
\ee
where the Gaussian kernel is defined as
\be 
\mathcal{D}_{\phi, \phi'}(t, s) = \frac{d\phi d\phi'}{2\pi g^2 J^2\sqrt{\det \mathcal{M}(t, s)}} \exp \left(-\frac{\bphi^T \mathcal{M}(t, s)^{-1}\bphi}{2 J^2 g^2} \right)
\ee
and
$$
\bM (t, s)=\left(
\begin{array}{cc}
C(t,t) & C(t,s)\\
C(s,t) & C(s,s)
\end{array}
\right)
$$
As noted above, in Sec. \ref{sec:s0-g0} and\ref{sec:sn0-g0}, chaotic motion is not time translation invariant, as trajectories are highly sensitive to their initial conditions; i.e. even small  initial perturbations can lead to vastly different outcomes as time progresses. 

In what follows, we examine the stability of stationary trajectories under small perturbations $\delta(t,s)$ that break time-translation invariance. Recalling that stationary solutions are characterized by a constant first moment $M(t) = M$ and time-translation invariant correlator $C(t, s) = C(t - s)$, we can write
\bea
C(t,s)&=&C(\tau)+\epsilon \delta(t,s)
\label{eq:C_p}
\\
C(t,t)&=&C(s,s)=C(0)
\label{eq:C0_p}
\\
M(t)&=&M
\label{eq:M_p}
\eea
where $\tau = t - s$, i.e. we consider a non-stationary regime where one-time quantities remain constant and two-time quantities break time-translation invariance ---much as happens in glassy regimes. 

Inserting Eqs.\eqref{eq:C_p}, \eqref{eq:C0_p} and \eqref{eq:M_p} in the equation for the two-time correlator $C(t,s)$, given by Eq.\eqref{eq:dot_Cts}, Taylor expanding the right hand side for small $\epsilon$ to linear order and using $C(\tau)=C(-\tau)$ and 
\be
F(C(\tau),C(0),C(0),C(-\tau),M,M)\equiv \Xi(C(\tau),C(0),M),
\ee
we get (i) to ${\mathcal O}(\epsilon^0)$ terms, Eq.\eqref{eq:dot_C};
and (ii), to ${\mathcal O}(\epsilon)$, the following equation for the deviations from stationarity 
\be
(1+\partial_t)(1+\partial_s)\delta(t,s) = \delta(t,s) \left.\frac{\partial \Xi(C,C(0),M)}{\partial C}\right 
\vert_{C(\tau)}.
\label{eq:dot_delta}
\ee
Following \cite{Helias2018}, we set 
$T=t+s$ and $\tau=t-s$ and denote $\delta(t,s)=K(T,\tau)$
to rewrite the above as 
\be
\left[(\partial_T+1)^2-\partial_\tau^2\right]K(T,\tau)=K(T,\tau)\, 
\Xi'(C(\tau),C(0),M)
\label{eq:PDE}
\ee
The above partial differential equation can be solved via separation of variables, i.e. by searching for a solution in the factorised form 
\be
K(T,\tau)=\psi(\tau)e^{\kappa T/2}
\ee
Inserting this ansatz into Eq.\eqref{eq:PDE}, a Schr\"odinger's equation for $\psi(\tau)$ is obtained
\begin{equation}
    [-\partial_\tau^2-V''(C(\tau)|C(0),M)]\psi(\tau)=
\left[1-(\frac{\kappa}{2}+1)^2\right]\psi(\tau)
\end{equation}
with quantum mechanical potential $-V''(C|C(0),M)$. 
As observed in \cite{Helias2018} there will be a set of allowed energies $E_0< E_1< E_2 \ldots$, with  $E_n=1-(\frac{\kappa_n}{2}+1)^2$, and 
associated eigenfunctions $\psi_n(\tau)$. 
A stationary solution $C(\tau)$ will be linearly stable if the perturbation $K(T,\tau)$ decays with $T$, i.e. if the largest value of $\kappa$, $\kappa_0=-1+\sqrt{1-E_0}$, is negative. This requires the ground state energy $E_0$ to be positive. 

It is also useful to note that, as pointed out in \cite{Helias2018}, 
Eq.\eqref{eq:dot_C} implies 
\begin{equation}
[-\partial_\tau^2-V''(C(\tau)|C(0),M)] \, \dot C(\tau)=0,
\label{eq:Cdot}
\end{equation}
hence non-fixed point steady-state solutions $\dot C(\tau)\neq 0$, correspond ---when they exist--- to eigenfunctions with energy $E=0$. 

As for an eigenfunction to exist, the energy must be greater or equal than the quantum-mechanical potential $-V''(C|C(0),M)$; if the latter is positive, all the eigenfunctions, including the ground state, have positive energy, hence the motion is stable.
\begin{enumerate}
    \item For $M\neq 0$, 
$V'(C|C(0),M)$ is strictly decreasing for any $-C(0)\leq C\leq C(0)$ and $0<C(0)\leq q$, hence $V''(C|C(0),M)<0$. This implies that for any $C(0)\leq q$, $E_n>0~\forall~n$ and the only allowed solution of Eq.\eqref{eq:Cdot} is the fixed-point solution $C(\tau)=q$ (non-trivial solutions with zero energy are not allowed). Furthermore, this must be stable, as $E_0>0$. 

\item Conversely, for $M=0$, the potential has at least one minimum, either at $C=0$ or $C=\pm \bar{C}(C(0))$, for any value of $C(0)\leq q$, hence Eq.\eqref{eq:Cdot} allows a non-trivial solution, for any $C(0)<q$, which is either a periodic orbit or the separatrix curve. These represent eigenfunctions with energy $E=0$. Periodic solutions have multiple nodes (as $\dot C$ vanishes at every turning point), whereas the separatrix curve has precisely one node in the noiseless case (where $\dot C(0)=0$) and no node in the noisy case (where $\dot C(0^+)=-\sigma^2$). As noted in \cite{Helias2018}, in the noiseless case $\dot C$ has at least one node, hence it cannot be the ground state, but it must correspond to one of the excited states $\psi_n(\tau)$ with $n\geq 1$, implying that the ground-state energy is $E_0<0$. 

Recalling that, in the noiseless case, steady-state solutions are physical only for $C(0)\leq q$, we have that {\it all} physical steady-state solutions are unstable in the noiseless case. 

On the other hand, in the presence of noise, solutions with $C(0)>q$ are allowed, {\it only} for $C(0)$ equating the separatrix value $C^\star_\sigma$, which solves Eq.\eqref{eq:Cstar-noise}. Since this is the only physical solution, it must be stable for $C_\sigma^*>q$. Conversely, any $C(0)<q$ leads to a physical solution. As for the noiseless case, such solutions are all unstable, except the separatrix curve, which is {\it marginally} stable (as it has no node for $\sigma\neq 0$). However, as any fluctuation  leads to an instability, chaotic behaviour is expected to emerge in single instances for $C_\sigma^*<q$. The critical line $C_\sigma^*=q$ is thus predicted to mark the transition to chaos. 
\end{enumerate}
 
\section{Phase Diagram for $\gamma \neq 0$}
\label{app:FP_diagram_gamma}

In this Appendix we derive the phase diagram for the fixed-point solutions of Eq.\eqref{eq:single-neuron}, when interactions are correlated. 
We note that for $g=0$, the solution of Eq.\eqref{eq:M-self-psi}-\eqref{eq:q-self-psi} is $M=0$, $q=0$. It is expected that $(M,q)=(0,0)$ remains stable for $g$ below a critical value $g_c$, where non-trivial solutions emerge. In this region,  Eq.\eqref{eq:x-self-psi} reduces to 
\be
x(\psi)=\tanh(\gamma g^2 J^2 x(\psi))
\ee
so that $x(\psi)$ is deterministic. We have $x(\psi)=0$ for $\gamma g^2 J^2<1$ --- either $\gamma>0$ and $(gJ)^{-1}>\sqrt{\gamma}$, or $\gamma<0$. 
For $\gamma>0$ and $(gJ)^{-1}<\sqrt{\gamma}$, two non-zero solutions emerge $\pm x^\star$, one positive and one negative. Since $x(\psi)=\pm x^\star$ implies $q\neq 0$, bifurcations in $x$ imply bifurcations in $q$, 
hence we assume that $x$ is small when $q$ is small.  
To locate $g_c$, we expand Eqs.(\ref{eq:M-self-psi})--(\ref{eq:q-self-psi}) for small $M$, $q$ and $x(\psi)$. Starting with Eq.\eqref{eq:M-self-psi}, we have 
\bea
M&=&\int D\psi [J_0g M+\gamma J^2 g^2 \chi x(\psi)
+gJ\sqrt{q}\psi]
\nonumber\\
&=&(J_0g+\gamma J^2 g^2 \chi) M
\label{eq:M-chi}
\eea
To make progress, we need an expression for $\chi$. Setting $\phi=gJ\sqrt{q}\psi$ we can write Eq.\eqref{eq:x-self-psi} as
\be
x(\phi)=\tanh [J_0g M+\gamma J^2 g^2 \chi x(\phi)+\phi]
\label{eq:x-self-phi}
\ee
Differentiating Eq. (\ref{eq:x-self-phi}) with respect to $\phi$ we obtain
\bea
\frac{\partial x}{\partial \phi}=\left[1-\tanh^2(gJ_0M+\gamma g^2J^2\chi x +\phi)\right]\Big(\gamma g^2 J^2 \chi \frac{\partial x}{\partial \phi}+1\Big),
\eea
and re-arranging  
\bea
\frac{\partial x}{\partial \phi}
\Big\{1-\gamma g^2 J^2 \chi
[1-\tanh^2(gJ_0M+\gamma g^2J^2\chi x +\phi)]
\Big\}
=1-\tanh^2(gJ_0M+\gamma g^2J^2\chi x +\phi),
\nonumber\\
\eea
so that,  writing in terms of $\psi$, we have 
\bea
\frac{1}{gJ\sqrt{q}}\frac{\partial x}{\partial \psi}
\Big\{1-\gamma g^2 J^2 \chi
[1-\tanh^2(gJ_0M+\gamma g^2J^2\chi x +gJ\sqrt{q}\psi)]
\Big\}
\nonumber\\
= 1-\tanh^2(gJ_0M+\gamma g^2J^2\chi x +gJ\sqrt{q}\psi)
\eea
Close to the transition line $M$, $q$ and $x$ are small. A leading-order expansion in these quantities gives
\be
\frac{1}{gJ\sqrt{q}}\frac{\partial x}{\partial \psi}
\Big\{1-\gamma g^2 J^2 \chi
\Big\}
=1.
\label{eq:dxdpsi}
\ee
Finally, averaging over $\psi$ and using Eq.(\ref{eq:chi-phi-psi}) we get 
\be
\chi(1-\gamma g^2J^2\chi)=1
\ee
The physical solution is 
\be
\chi=\frac{1-\sqrt{1-4\gamma g^2J^2}}{2\gamma g^2 J^2}
\label{eq:chi-physical}
\ee
(as the other diverges when $g\to 0$). 
Substituting in Eq.(\ref{eq:M-chi}), we have that $M$ bifurcates at 
\be
gJ_0+\gamma g^2 J^2 \frac{1-\sqrt{1-4\gamma g^2J^2}}{2\gamma g^2 J^2}=1
\ee
which gives 
\be
\frac{1}{gJ}=\frac{J_0}{J}+\gamma\frac{J}{J_0}
\label{eq:out-asym}
\ee
Similarly, expanding the equation for 
$q$, Eq.\eqref{eq:q-self-psi}, and setting $M=0$ we obtain
\bea
q&=&\int \D\psi[J_0g M+\gamma J^2 g^2 \chi x(\psi)
+gJ\sqrt{q}\psi]^2
=\gamma^2 J^4 g^4 \chi^2q
+g^2J^2q+2\gamma (gJ)^3\chi\sqrt{q}\int \D\psi x(\psi)\psi
\nonumber\\
&=&
(\gamma^2 J^4 g^4 \chi^2
+g^2J^2+2\gamma (gJ)^4\chi^2) q
\nonumber\\
\label{eq:use-parts}
\eea
where we have integrated by parts
\bea
\int\D\psi x(\psi)\psi&=&\int \frac{d\psi}{\sqrt{2\pi}}\Big(-\frac{\partial}{\partial\phi}e^{-\psi^2/2}\Big)x(\psi)
=\int \frac{d\psi}{\sqrt{2\pi}}e^{-\psi^2/2}\frac{\partial }{\partial \psi}x(\psi)
=\left\langle \frac{\partial x}{\partial \psi}\right\rangle=gJ\sqrt{q}\chi.
\label{eq:xpsi}
\eea
Substituting Eq.(\ref{eq:chi-physical}) in Eq.(\ref{eq:use-parts}) we see that 
bifurcations in $q$ occur at
\be
1=\frac{\gamma+2}{2\gamma}(1-\sqrt{1-4\gamma g^2J^2}-2\gamma g^2J^2)+g^2J^2
\ee
which can be simplified to 
\be
\frac{2}{g^4J^4}-(2\gamma^2+3\gamma+2)\frac{1}{g^2J^2}-\gamma(\gamma+1)^2=0.
\ee
As the physical solution must be positive, it follows that
\be
\frac{1}{gJ}=\frac{1}{2}\sqrt{(2\gamma^2+3\gamma+2)\pm\sqrt{(2\gamma^2+3\gamma+2)^2+8\gamma(\gamma+1)^2}}.
\ee
With little algebra, the expression above can be simplified to 
\be
\frac{1}{gJ}=\frac{1}{2}\sqrt{2+\gamma(2\gamma+3)\pm|2\gamma^2+5\gamma+2|}\equiv a(\gamma).
\ee
As $2\gamma^2+5\gamma+2\geq 0$ for $\gamma \geq -1/2$, we have 
\be
a(\gamma)=\left\{
\begin{array}{cc}
\frac{1}{2}\sqrt{2+\gamma(2\gamma+3)\pm(2\gamma^2+5\gamma+2)} & \quad\quad \gamma \geq -1/2 \\
\frac{1}{2}\sqrt{2+\gamma(2\gamma+3)\mp (2\gamma^2+5\gamma+2)} & \quad\quad\gamma<-1/2
\end{array}
\right.
\ee
hence there are two possible solutions, $a_+(\gamma)=1+\gamma$ and $a_-(\gamma)=\sqrt{-\gamma/2}$, the latter existing only for $\gamma<0$. 
As for $\gamma=-1$, all the eigenvalues of the interaction matrix $\bJ$ are purely imaginary, no (non-trivial) fixed-point solution are expected to bifurcate from $\bx=0$, as bifurcating solutions must be purely oscillatory. Hence, it is expected that $a(-1)=0$. 
This suggests that the solution $a_+(\gamma)$ is the physical one for any $\gamma\in [-1,1]$. 
Combining Eq.\eqref{eq:out-asym} 
with the result $a(\gamma)=1+\gamma$, 
we obtain the critical line where the paramagnetic phase becomes unstable given in Eq.\eqref{eq:critical-line-gamma}.

\section{Numerical procedure}
\subsection{Computation of trajectories}
In simulations, trajectories of Eq.(\ref{eq:dynamics-def}) are computed numerically by a simple Midpoint Runge-Kutta method (see, for instance, \cite{toral_stochastic_2014}). The time step is fixed $h = 0.1$, except for the calculation of the LLE for which $h = 0.01$ because of the numerical sensitivity of the Bennetin-Wolf procedure. In order to avoid the transient regime, trajectories are integrated up to $t_{max} = 2000$ units of time. The initial step $\bx_0$ and the matrix $\mathbf{J}$ are selected randomly for each instance of the simulation. In particular, each pair $(J_{ij}, J_{ji})$, for $1 \leq i < j \leq N$, is generated following a multivariate Gaussian distribution with average $\bmu$ and covariance matrix ${\bSigma}$, as defined in \eqref{eq:mu-Sigma}, respectively.
The diagonal elements, $J_{ii}$, are each obtained from a Gaussian distribution with average $J_0/N$ and variance $J^2/N$.

\subsection{Shape of the potential $V(C | C(0), M)$}
In order to numerically compute $V(C |C(0), M)$, we simplify the function $\Xi(C, C(0), M)$, defined in Eq.(\ref{eq:Xi}),
as follows.
The Gaussian kernel, Eq.(\ref{eq:Gaussian_kernel}), is divided into two parts using the conditional probability formula of a multivariate-Gaussian distribution,  i.e. the probability of $\phi$ respect to fixed $\phi'$. In these terms, one can write 
\begin{equation}
\begin{cases}
      \Phi(\phi, \phi') & =  g \left[ J (\sqrt{C_0 - C^2 / C_0} \,\phi + C / \sqrt{C_0}\, \phi') + J_0 M \right] \\ 
      \Psi(\phi) &=  g \left [J \sqrt{C_0} \, \phi' + J_0 M\right]
\end{cases}
\end{equation}
then, by means of the Fubini's theorem, function $\Xi(C, C(0), M)$ becomes 
\begin{equation}
    \Xi = \int \frac{d \phi}{\sqrt{2 \pi}} e^{- \phi^2/2} \, \tanh(\Psi(\phi)) \, \left\{\int \frac{d \phi'}{\sqrt{2 \pi}} e^{- \phi'^2/2} \, \tanh(\Phi(\phi, \phi')) \right \}.
\end{equation}
With this simple form, it is straightforward to compute $V(C |C(0), M)$ directly by solving numerically the iterated integrals. 

\subsection{Computation of the largest Lyapunov exponent}
\label{app:LLE}

In order to compute the LLE we use the Bennetin-Wolf algorithm \cite{eckmann_ergodic_1985, pikovsky_lyapunov_2016}, that works as follows. Consider a dynamical system described by the vector $\bx(t) \in \mathbb{R}^N$ at time $t \geq 0$, which evolves following a differentiable dynamical system given by $\dot \bx(t) = \bF(\bx(t))$. In order to integrate it numerically, we construct the map 
\begin{equation*}
    \bx_{i + 1} = \bff(\bx_i)
\end{equation*}
where $\bff$ is the discretization of the vector function $\bF$ (which depends on the integration algorithm used) and $\{\bx_i: i = 0, 1, 2, ...\}$ defines the discrete trajectory of the system. We use a simple integration algorithm, i.e. the Euler's method: the vector function $\bf$ is defined as the linear evolution of state $\bx$ in a time-step $h$, 
\begin{equation*}
    \bff(\bx) \equiv \bx + \bF(\bx) \, h. 
\end{equation*}
Upon defining the matrix $\bT(\bx) = \mathcal{J}_\bx \mathbf{f}$, where $\mathcal{J}_\bx \mathbf{f}$ is the $N\times N$ Jacobian matrix of the vector function $\bff$ at state $\bx$, we have $\bT(\bx_{i + 1}) = \mathbb{\bI} + h\, \mathcal{J}_{\bx_{i + 1}} \bF$. Writing 
\be
\bT_{\bx_0}^n = \bT(\bx_{n-1}) \dots \bT(\bx_1) \bT(\bx_0),
\ee
as a matrix product, the LLE can be computed as the limit
\begin{equation}
    \Lambda = \lim_{n \longrightarrow \infty} \frac{1}{n} \log \Vert \bT_{\bx_0}^n \bu \Vert 
\end{equation}
for a given vector $\bu$, where  $\Vert \bv \Vert$ denotes the norm of the vector $\bv$. To implement this method computationally, the Bennetin-Wolf correction \cite{benettin_lyapunov_1980, wolf_determining_1985} needs to be used to avoid the divergence of $\bT_{\bx_0}^n \bu$ for large $n$. The correction consists in normalizing this vector at each step. 
Hence, fixing a starting point $\bx_0$, and defining an initial vector $\bu_0 = [1, ..., 1] / \sqrt{N}$, the algorithm is described as follows: for $i = 1, \ldots, T$, 
\begin{enumerate}
    \item Obtain the new state  at $i+1$,  
        \begin{equation*}
        \begin{array}{ll}
        t_{i + 1} = & t_i + h \\
        \bx_{i + 1} = & \bx_{i} + h \, \bF(\bx_i)  \\
        \tilde \bu_{i + 1} = & \bT(\bx_{i+1}) \, \bu_{i }
        \end{array}
        \end{equation*}
        \item Normalize the perturbation vector
        \begin{equation*}        
         \bu_{i + 1} = \tilde \bu_{i + 1} / \Vert \tilde \bu_{i + 1} \Vert
        \end{equation*}
    \item Compute the LLE at time-step $i + 1$, 
        \begin{equation*}
            L_{i + 1} = L_i + \log(\Vert\tilde \bu_{i + 1}\Vert) / t_{i + 1}
        \end{equation*}        
\end{enumerate}
The algorithm gives a set $\{\L_i : i = 0, ..., T\}$ of exponents which, in the limit, converge to the LLE of the system, $\Lambda$.

\begin{figure}[tbh]
    \centering
    \includegraphics[width=1 \linewidth]{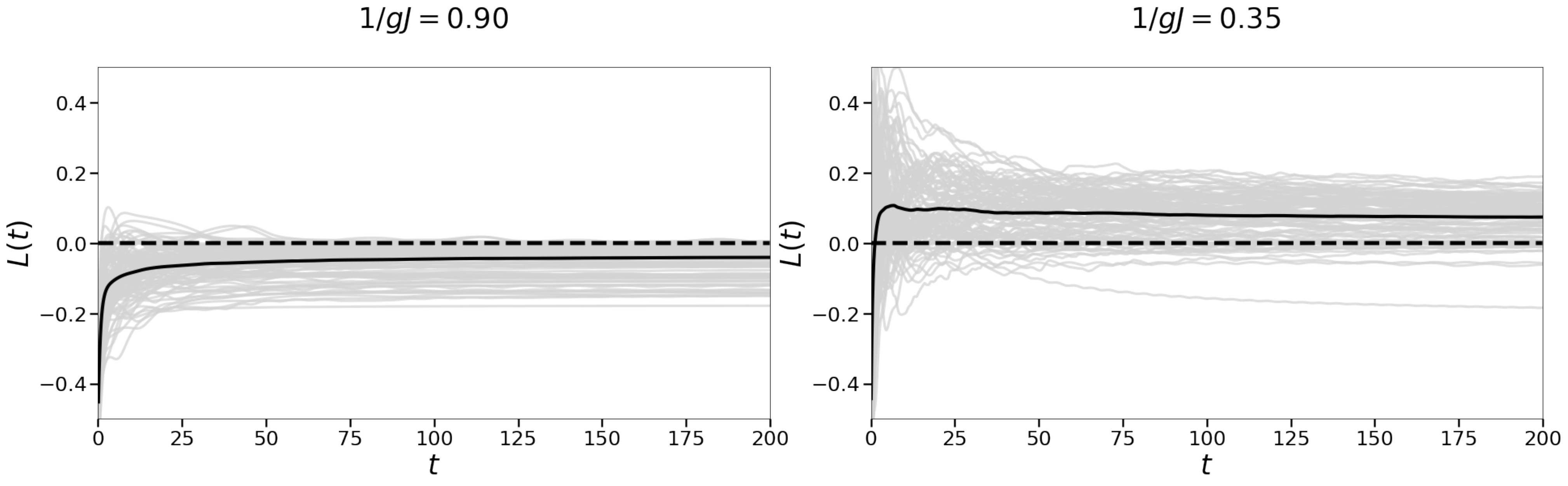}
    \caption{Computation of largest Lyapunov exponent for $S = 100$ different realizations of a system with $N = 100$, inside the spin-glass regime, above (left panel) and below the onset of chaos (right panel). Each realization is plotted as a grey, solid line, meanwhile the black, solid line is the average over realizations for each time step. }  
    \label{fig:Lyapunov_1}
\end{figure}

Consider now a stochastic differential equation, $\dot \bx(t) = \bF(\bx(t)) + \bxi(t)$ such that $\bxi(t)$ is a (multi-component) white-noise process with zero mean and variance equal to $2 \sigma^2$. Hence, the previous algorithm should be modified to introduce the stochastic integration: at step 1., the integration of the trajectory should be replaced by an Euler-Maruyama integration step \cite{toral_stochastic_2014}, 
\be
\bx_{i + 1} = \bx_i + \bF(\bx_i) h + \sqrt{2 h \sigma^2} \bu_i
\ee
where $\bu_i$, for $i = 1, 2, \ldots $, is a set of i.i.d. random Gaussian variables with zero mean  and unit variance. 

For the dynamical system given by Eq.(\ref{eq:dynamics-def}), it is sufficient to fix $h = 0.01$, and integrate the system up to $t_{max} = 200$. As an example, in Fig.\ref{fig:Lyapunov_1} we show $S = 100$ realizations of the algorithm for a system of size $N = 100$ (grey, solid line) and $\sigma^2 = 0$, each of them initialized with random conditions, $\bx_0$ and $\mathbf{J}$. The system is set at $J_0/J = 0.5$ (i.e. inside the spin-glass phase). We show results for $1/gJ = 0.9$ (left panel) and $1/gJ = 0.35$ (right panel). 
 The solid, black line indicates the average over different realizations at each time step, showing that the mean activity is stable (negative LLE) for $1/gJ = 0.9$ and unstable (positive LLE) for $1/gJ = 0.35$. In the latter case, there exists a large amount of unstable trajectories, such that  $L_i$ is positive, and some stable trajectories, such that $L_i$ negative, e.g. fixed points. 

\begin{figure}[tbh]
    \centering
    \includegraphics[width=0.4 \linewidth]{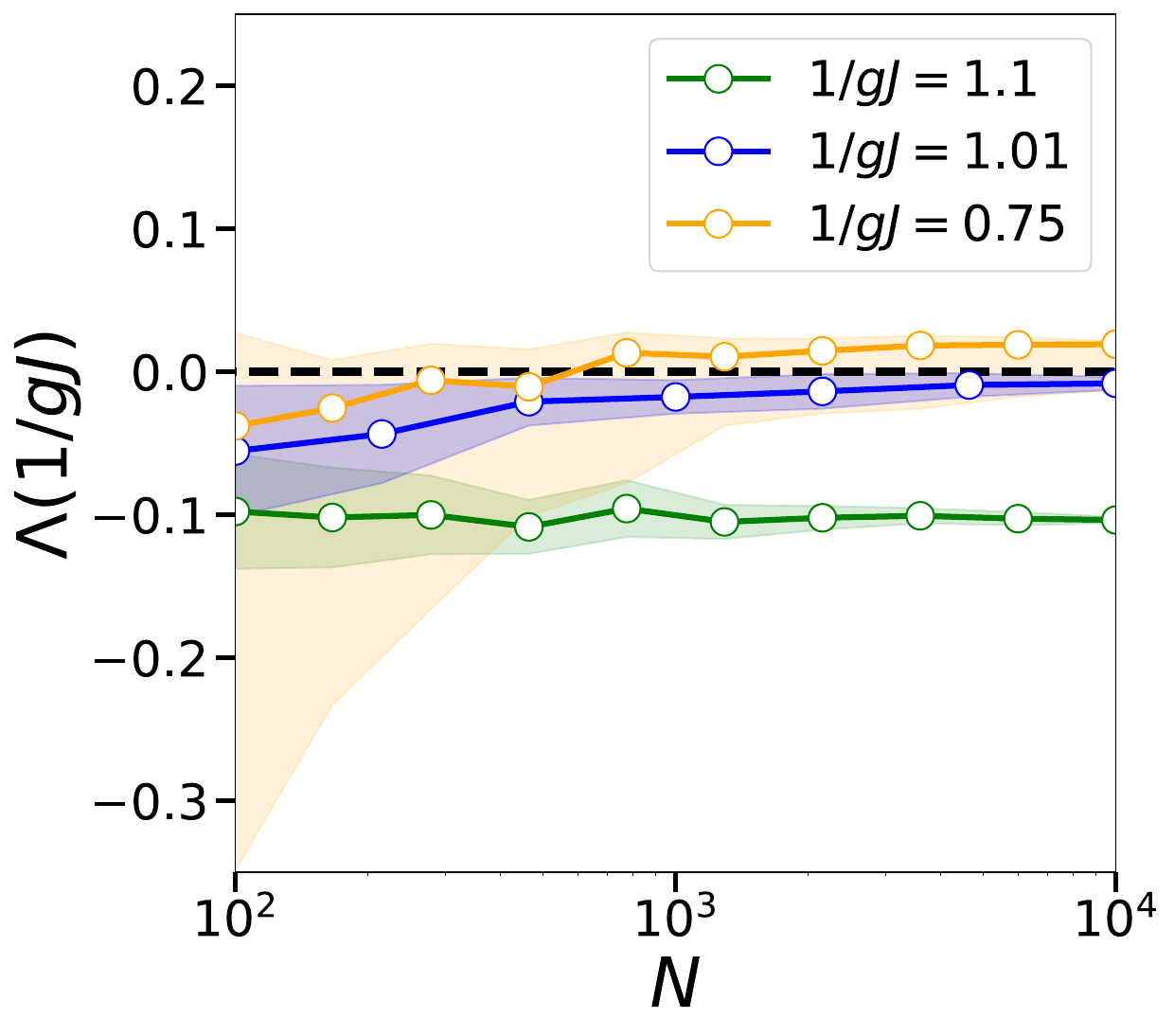} 
    \caption{Averaged largest Lyapunov exponent, $\Lambda$, as a function of system size $N$ for a dynamical system with $\sigma^2 = 0$, averaged over $S = 20$ realizations. For $J_0/J = 0.5$, the averaged LLE  is computed for three different values of $1/gJ$ (solid lines), as shown in the legend; the standard deviations are shown as shaded areas. Parameter values are as in Fig.\ref{fig:Lyapunov_1}, i.e. $h = 0.01$ and $t_{max} = 200$. }
    \label{fig:Lyapunov_2}
\end{figure}

In order to study the convergence, we compute the LLE as a function of the system size $N$. In  Fig.\ref{fig:Lyapunov_2} we plot, as an example, the LLE $\Lambda$, defined as the asymptotic value of $L(t)$ (i.e. the value at $t = 200$), averaged over $S = 20$ realizations, for the particular case of $\sigma = 0$. Fixing $J_0/J = 0.5$, we study three different values of $1/gJ$ in order to characterize the stability of the system in the limit of large system size: $1/gJ = 1.1$ (solid green line) corresponding to the paramagnetic phase; $1/gJ = 1.01$ (solid blue line) falling close to the critical line separating the spin-glass and the paramagnetic region; $1/gJ = 0.75$ (solid orange line), lying in the spin-glass region, away from the critical line. Colored area shows the standard deviation of $\Lambda$ for each value of $N$. For $1/gJ > 1$ (blue and green lines), the LLE is typically negative, indicating the existence of stable, fixed point trajectories. In this region, the standard deviation of the LLE is smaller compared to the $1/gJ <1$ case (orange line). The reason behind this is that, above this critical value of $1/gJ$, the system converges towards fixed point solutions; while below it, finite size effects in the network encourage the appearance of many different trajectories: chaotic, limit cycles and even, occasionally, fixed points, all of it contributing to increasing the standard deviation of the LLE. In practice, for large enough systems (i.e. $N > 1000$) the LLE stabilizes considerably and the standard deviation reduces, as it can be seen in Fig.\ref{fig:non_fixed_points}.

\end{appendix}

\bibliography{Types.bib}

\end{document}